\documentclass[11pt]{article}

\usepackage{epsf}
\def\1ad{\mbox{\normalsize $^1$}}
\def\2ad{\mbox{\normalsize $^2$}}
\def\3ad{\mbox{\normalsize $^3$}}
\def\4ad{\mbox{\normalsize $^4$}}
\def\5ad{\mbox{\normalsize $^5$}}
\def\6ad{\mbox{\normalsize $^6$}}
\def\7ad{\mbox{\normalsize $^7$}}
\def\8ad{\mbox{\normalsize $^8$}}

\def\dj{\hbox{d\kern-0.347em \vrule width 0.3em height 1.252ex depth
-1.21ex \kern 0.051em}}

\def\shalf{\mbox{${1\over 2}\,$}}

\def\Tr{{\rm Tr\,}}

\def\ket{\rangle}
\def\bra{\langle}

\def\pt{\partial}

\def\Tr{\mbox{Tr}}
\def\bra{\langle} 
\def\ket{\rangle} 
\def\lim{\mbox{{\bf L}} }

\newcommand{\plb}[3]{{{Phys.~Lett.~B}~{\bf #1},  #2 (#3)}}
\newcommand{\npb}[3]{{{Nucl.~Phys.~B}~{\bf #1},  #2 (#3)}}
\newcommand{\prd}[3]{{{Phys.~Rev.~D}~{\bf #1},  #2 (#3)}}

\newcommand{\ijmpa}[3]{{{Int.~J.~Mod.~Phys.~A}~{\bf #1},  #2 (#3)}}
\newcommand{\prl}[3]{{{Phys.~Rev.~Lett.}~{\bf #1},  #2 (#3)}}

\newcommand{\hepth}[1]{[hep-th/#1]}

\newcommand{\ben}{\begin{equation*}}
\newcommand{\een}{\end{equation*}}
\newcommand{\ba}{\begin{eqnarray}}
\newcommand{\ea}{\end{eqnarray}}
\newcommand{\ban}{\begin{eqnarray*}}
\newcommand{\ean}{\end{eqnarray*}}
\newcommand{\brr}{\begin{array}}
\newcommand{\err}{\end{array}}
\newcommand{\bc}{\begin{center}}
\newcommand{\ec}{\end{center}}

\def\be{\begin{equation}}
\def\ee{\end{equation}}
\def\ve{\varepsilon} 

\begin{document}

{\large
\newcommand{\sheptitle}
{Touring the Hagedorn Ridge}
\newcommand{\shepauthora}
{{\sc
 J.L.F.~Barb\'on}
}

\newcommand{\shepaddressa}
{\sl
Department of Physics, Theory Division, CERN \\
 CH-1211 Geneva 23, Switzerland \\
{\tt barbon@cern.ch}}

\newcommand{\shepauthorb}
{\sc
E.~Rabinovici}

\newcommand{\shepaddressb}
{\sl
Racah Institute of Physics, The Hebrew University \\ Jerusalem 91904, Israel
 \\
{\tt eliezer@vms.huji.ac.il}}

\newcommand{\shepabstract}
{ We review aspects of  the Hagedorn regime
in critical string theories, from  basic facts about
the ideal gas approximation to the proposal of a global picture 
inspired by general ideas of holography. It was    
suggested  
that the condensation of thermal winding modes triggers a first-order phase
transition. We propose, by an Euclidean analogue of the string/black hole 
correspondence principle, that the transition is actually related 
to a 
topology change in  spacetime. 
Similar  
phase transitions 
induced by unstable winding modes can be studied in toy models. There,
using  T-duality
of supersymmetric cycles, one can identify 
a topology change of the Gregory--Laflamme type,
which we associate with   large-$N$ phase transitions of Yang--Mills
theories on tori. This essay is dedicated to the memory of Ian Kogan. 
}

\begin{titlepage}
\begin{flushright}
{CERN-PH-TH/2004-141 \\
{\tt hep-th/0407236}}

\end{flushright}
\vspace{0.5in}
\begin{center}
{\Large{\bf \sheptitle}}
\bigskip\bigskip \\ \shepauthora \\ \mbox{} \\ {\it \shepaddressa} \\
\vspace{0.3in}
\bigskip\bigskip  \shepauthorb \\ \mbox{} \\ {\it \shepaddressb} \\
\vspace{0.3in}

{\bf Abstract} \bigskip \end{center} \setcounter{page}{0}
 \shepabstract
\vspace{0.3in}
\begin{flushleft}
CERN-PH-TH/2004-144\\
\today
\end{flushleft}

\end{titlepage}
}

\newpage

\setcounter{equation}{0}

\section{Introduction}

\noindent

Perturbative string theory manifests several  bounds. One of them
is a seemingly upper bound on the allowed value of the temperature
of a  string gas --the Hagedorn temperature \cite{hag, bunch}. Bounds are
there to be understood and challenged. Ian has challenged this one and
has taken a peek beyond it in his seminal 1987 work \cite{ian}. In this
essay we return to discuss the possible instabilities and tachyons emerging
near the Hagedorn temperature. We do it equipped with tools that Ian has
forged in collaboration with us \cite{thresholds, usabel, golfand}.

The spectrum of a finite-tension critical string in perturbation theory
 has two
universal components: the first is familiar from point particles, it consists
of a finite set of
massless modes that include
gauge fields (open strings) and gravitons (closed strings); the second is
of a stringy nature and consists of an exponential
degeneracy of states at a given high energy. It is the second component that
gives rise to a limiting temperature.   For a single string of
energy $\varepsilon$ the density of states grows very roughly as  $
\omega(\varepsilon) \sim \exp(\beta_s\,\varepsilon)$,
where $\beta_s \sim \ell_s$ is of the order of the string length scale (we set
$\ell_s =1$ in the following and measure all dimensionful quantities in string units).
Its entropy is
\begin{equation}\label{hagen}
S(\varepsilon) = \log\,\omega(\varepsilon) \sim \beta_s\,\varepsilon
\,,\end{equation}
and its effective temperature is obtained through the relation
\begin{equation}\label{hagent}
{1\over T} = {\pt S \over \pt \varepsilon} \sim \beta_s
\,.\end{equation}
We see that this temperature is approximately bounded by the constant
$T_s =1/\beta_s$, called the  Hagedorn temperature. Essentially all the
energy pumped into the system is utilized to create the large number of
new particles becoming available as the energy increases, instead of
increasing the energy of the particles already present at lower energy. Thus
keeping the temperature fixed.

 Field-theoretical entropies, such as those of each of the massless modes,
scale in $d$ spatial dimensions as $E^{d /( d+1)}$. Hence, the highly excited
strings dominate any thermal state beyond string-scale  energy densities.
Since the entropy is approximately independent of
the energy,  the resulting specific heat seems infinite. In fact it turns out
that small corrections to
(\ref{hagen}) can drive the system either into a stable phase of
positive specific heat or to an unstable one of negative specific heat.

A limiting temperature was detected in several types of systems. 
Historically, it was first observed in the dual theory of hadrons and
 the first  physical interpretation of  Hagedorn's  ``limiting
temperature" was offered in the QCD theory of hadrons. The answer in  QCD
is certainly dramatic: it was suggested that instead of being an actual
limiting temperature its presence suggests a change in the relevant
degrees of freedom in terms of which the system is relatively simple. A change
resulting in a phase transition from composite objects to their constituents
\cite{cab, thorn}.
The Hagedorn temperature in hadronic systems is related to a ``deconfinement" transition
in which the hadrons liberate their quark-gluon constituents. Once the degrees of freedom
are expressed in terms of the field theory of quarks and gluons there is no bound
on the temperature; it  can be raised indefinitely.
This  QCD analogy has been a recurrent theme when thinking about
 the ``fundamental
strings" of quantum gravity and their possible ``true constituents". This was the
problem Ian tackled.
 The advent of the AdS/CFT correspondence \cite{adscft} has made
specific models amendable to a non-perturbative
analysis, allowing to reformulate (and sometimes answer) these
old questions. By a twist, 
 a physical picture arises that resembles QCD very closely,
and  links  to gravity and ten-dimensional physics
by the magic of holography \cite{holo}.

 We start this essay in section 2  by introducing, in order of appearance, 
the cast of
degrees of freedom relevant for each appropriate energy scale. 
 We paint with large brush strokes the dependence of the
temperature on the energy of each set of degrees of freedom, and we find that
the  system crosses
the limiting temperature protected by a black hole armor. 
In section 3 we review
the more detailed high energy behavior of the spectra
of various types of strings. This is illustrated using a simple and
useful geometrical picture:
 that of random walks. Particular attention is paid to
the dependence of the spectra on the large-distance properties of the background
geometry. In sections 4 and 5
 an Euclidean picture of the physics around the Hagedorn
temperature is discussed. The physical significance of the thermal tachyon discovered
by Ian is addressed, and its stringy features  are underscored. The conclusion
that this tachyon seems to be more of a book keeping device than a physical particle
is deconstructed.  Instead, a very physical Euclidean picture emerges; the
transition monitors a change in the topology of spacetime enforced by
the nucleation of black holes. The manifestation  of the tachyon is a specialization
of the mechanism envisaged by Ian. We finish in section 6 with a discussion of a toy
model in which T-duality goes a long way towards  solving a similar problem, involving
a dynamical topology change.

This review is centered around the system of  ten-dimensional critical strings
at finite temperature. Generalizations of these questions to more
exotic backgrounds of particular interest, such as LST \cite{lst} and 
PP waves \cite{bmn}, have emerged recently. We will not discuss these issues
here, and the interested reader may consult
\cite{kutsaha} and \cite{bmnt} for results and 
lists of references.

\section{String thermodynamics: the big picture }

\noindent

The first two characters in the cast of constituent ingredients, 
out of which the gravity cocktail is composed,  
are  the massless modes
with field-theoretical entropy of order $E^{d/ 
(d+1)}$ and the highly-excited
strings with entropy proportional to their energy. Here we are assuming that the
spatial volume, $V$, is finite, 
 the string coupling is sufficiently small, $g_s \ll 1$, and 
 the local spacetime
 geometry is approximately flat ${\bf R}^{d+1}$ over the length scales of the
box of volume $V=L^d$. The simplest string background with these properties is
a spatial  toroidal compactification with $d$ dimensions of size $L$,  
 $9-d$ dimensions  of string-scale size, and a very small string coupling, so that
 we can measure energies with respect to the flat time coordinate, at least to the
 extent that gravitational back reaction can be neglected.  Maximizing the entropy
at a given total energy among these two components, one finds the temperature
dependence on the energy,  $T(E) = (\pt S /\pt E)^{-1}$. This is shown in figure 1.

\begin{figure}[ht]
\centerline{\epsfxsize=1.9in\epsfbox{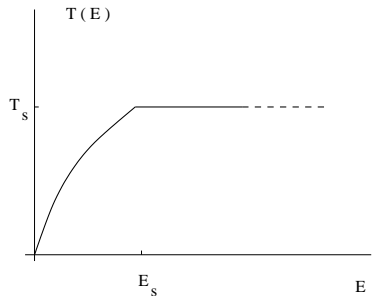}}
\caption{\sl The microcanonical temperature function $T(E) = (\pt S /\pt E )^{-1}$
for a string gas in the two-component approximation. For energies
$0<E<E_s$ below the string-scale threshold,
 with $E_s = \rho_s V$, and $\rho_s = O(1)$ in string units, the temperature
grows as $T\sim E^{1/ (d+1)}$, dominated by the massless modes. It gets  
saturated at $T\approx T_s$ by the highly excited strings. The dotted line
represents the sensitivity to small interaction effects that can perturb
the Hagedorn plateau either way, into a regime of positive specific heat
(with $dT(E)/dE >0$) or a negative one.
 }
\end{figure}

In this approximation, a temperature
plateau seems to emerge. Moreover,   the ``Hagedorn band"
is not sensitive to the number distribution of strings, i.e. the result
is the same whether we assume that a single string carries all the
available energy, or rather the energy is distributed among various strings (provided
all of them carry enough excitation energy to be in the Hagedorn regime).

\begin{figure}[ht]
\centerline{\epsfxsize=2.9in\epsfbox{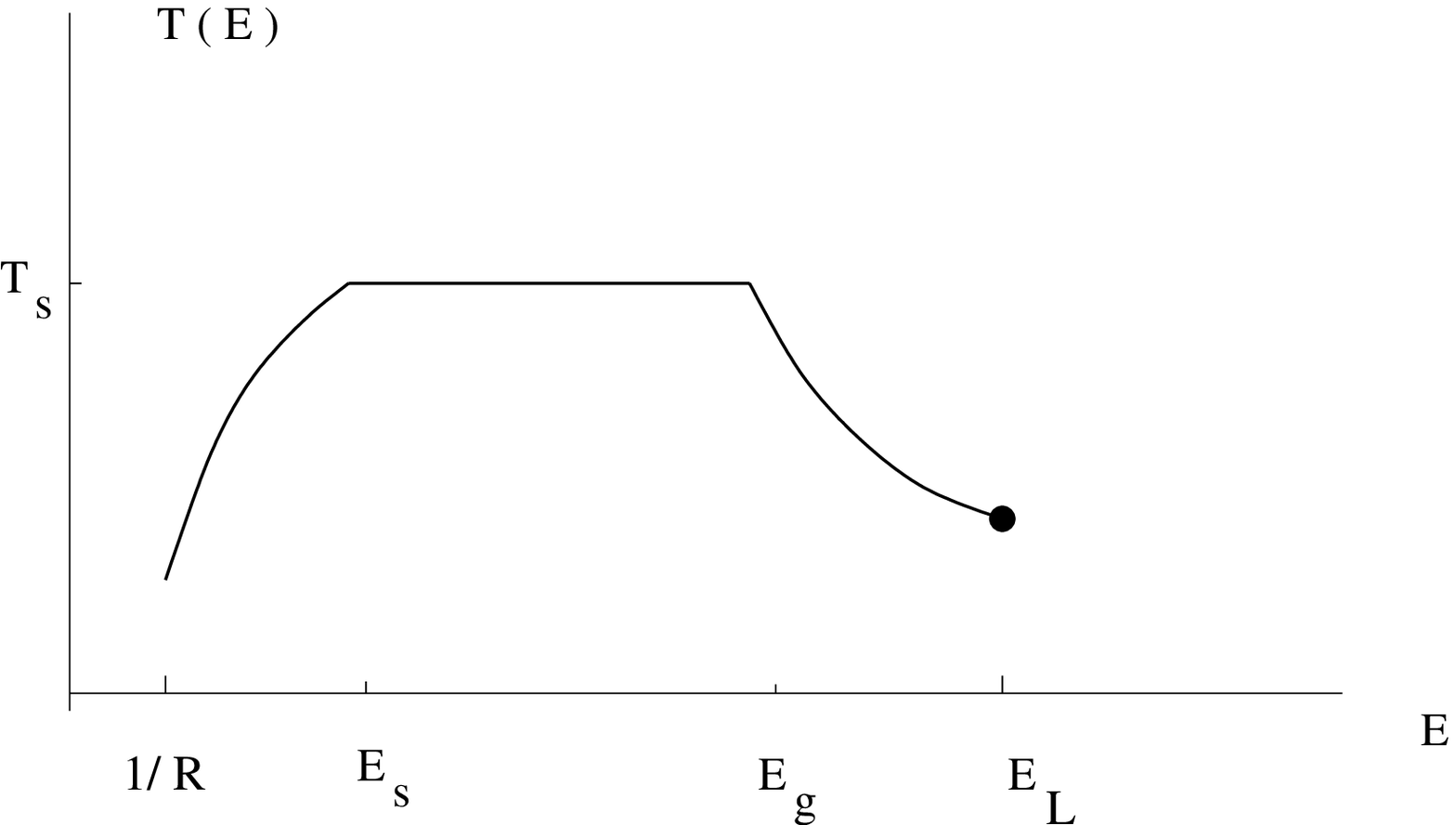}}
\caption{\sl The temperature function in the three-component approximation. 
Black hole dominance of the density of states at very high energies
implies a phase of negative specific heat, starting at $E_g \sim 1/g_s^2$
 (in string units),
corresponding to the nucleation of small black holes of size $\ell_s$, that
subsequently grow as the temperature drops. The Jeans energy $E_L \sim
L^{d-2} /g_s^2$ represents the limit beyond which back-reaction effects
cannot be neglected on the scale $L$ at the finite-volume box. We have also
included the low-energy cutoff at energies of order $1/L$ because the
standard scaling $T\sim E^{1/(d+1)}$  only applies for energies above the
gap of finite-volume excitations.  
 }
\end{figure}

Corrections to this rough picture depend on the details of the interactions.
This is a characteristically difficult problem, since we are in a regime
with large breaking of supersymmetry where no standard approximations are
available. For the case $d>2$, a qualitative
 picture confirms what is expected from 
the  principle of ``asymptotic
darkness" \cite{banks}, which states that
 black holes dominate the extreme
high-energy regime of theories that incorporate gravity 
(in three spacetime dimensions or less, ``asymptotic darkness" arguments require
special care, since  localized energy sources affect
asymptotic conditions for the vacuum). String theory is
no exception in this regard, having its own correspondence principle \cite{hp, corr, cor}.
 In particular, the entropy of Schwarzschild black
holes in $d$ dimensions scales as
$$
S\sim E \,(g_s^2 \,E)^{1 \over d-2}
\;,$$
 and eventually
dominates over the Hagedorn degeneracy for $g_s^2 E >1$. At this point the
Hagedorn plateau must end  and drop to lower temperatures $T(E) \sim E^{1/
(2-d)}$, which is a phase of negative specific heat for $d>2$.
 In fact, this phase
cannot be continued to arbitrarily high energies because the black hole
eventually grows to the size of the box. This threshold coincides with
the Jeans length entering inside the box,
and corresponds to an energy
\begin{equation}\label{jeans}
E_L \sim {L^{d-2} \over g_s^2} \;.
\end{equation}
The bound $E<E_L$ implies that no thermodynamic limit (large volume with
constant energy density) is possible in these
systems, since $E_L /V \rightarrow 0$.

An exit out of this tight corner is provided by an appropriate infrared
regularization.
An interesting model for a ``box" is obtained by  replacing flat space by
Anti-de Sitter space, ${\rm AdS}_{d+1}$ (c.f. \cite{HP}).
 Consistent string backgrounds exist with AdS factors,
the simplest one being the extensively studied ${\rm AdS}_5 \times {\bf S}^5$ background
of type IIB strings.
 In such a space,
the gravitational redshift effectively confines finite-energy excitations
within a distance of order $R$, the  radius of negative curvature.
Black holes larger than $R$ exist but have new features,  the most
important being their positive specific heat. The
Bekenstein--Hawking entropy of these black holes scales as
\begin{equation}
S\sim (E\,R)^{d-1 \over d}
\,,
\end{equation}
exactly like a conformal field theory (CFT) in a $(d-1)$-dimensional box of
size $R$. One may attempt to consider this as an 
embodiment of ``asymptotic darkness": 
a string system whose black holes have only a field-theory type entropy will
not be able to posses more than field-theory entropy at high energies.
Assuming that the cast of characters is complete, we can draw a global phase diagram,
as in figure 3, that sums up our general  knowledge of the Hagedorn regime in
critical string theory \cite{usabel, adshag, beken, magoo}.  
 
\begin{figure}[ht]
\centerline{\epsfxsize=2.9in\epsfbox{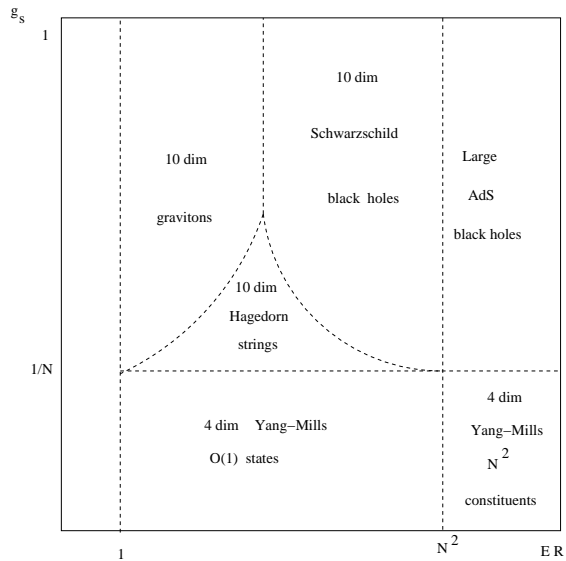}}
\caption{\sl The phase diagram of quantum gravity in ${\rm AdS}_5 \times
{\bf S}^5$ with $N$ units of Ramond--Ramond flux,
 according to the AdS/CFT correspondence. A ten-dimensional
weakly-curved description only arises for $R/\ell_s \sim (g_s N)^{1/4} \gg 1$.
In this case, the Hagedorn regime is bounded by black hole and massless
graviton phases. It only exists for sufficiently weak string coupling, 
$g_s \ll  
N^{-9/17}$. At very weak couplings, $g_s < 1/N$, one must use four-dimensional
descriptions based on the Yang--Mills degrees of freedom, whereas at strong
coupling, $g_s >1$, the diagram mirrors itself by the action of S-duality. 
 }
\end{figure}

We see that the standard ten-dimensional  Hagedorn phase
is  always a bounded transient, to be exited at high energies
by a black hole phase. If the string coupling is too large, for a given ratio $R/\ell_s$, 
the Hagedorn plateau is not even present, as the black holes form and
 drop the temperature
before  reaching
 $T_s$. On the other hand, at very weak coupling, $g_s < 1/N$, one
has $R<\ell_s$ and the ten-dimensional spacetime is strongly curved. In this case it
is better to describe the system in dual Yang--Mills variables.
 The large black hole goes over the quark-gluon plasma phase, whereas the
ten-dimensional Hagedorn regime goes over the four-dimensional glueball
regime (thus we are back to the dual models). In this situation, the fate of the
Hagedorn plateau must be analyzed in Yang--Mills perturbation theory  \cite{aha}
(c.f. figure 5). 
  
We can draw the microcanonical temperature function by cutting the phase diagram
at fixed string coupling, within the limits $N^{-1} < g_s < N^{-9/17}$. 
The new branch of AdS black holes allows to extend the function $T(E)$
beyond the Jeans energy to indefinite energies \cite{w, adshag}. The temperature also
rises indefinitely, as  shown
in figure 4. Actually one can read off figure 4 the phase structure
of the system as a function of the temperature. It has a first-order phase transition
with a latent heat of $O(g_s^{-2})$ and a critical  temperature $T_c \sim 1/R$.  In this
case, a large black hole nucleates much before the temperature can reach
the Hagedorn domain. Thus, the Hagedorn regime is a superheated phase
which is either unstable or weakly metastable to decay into the large
black hole phase, which engulfs all the space occupied by the hot string gas.

\begin{figure}[ht]
\centerline{\epsfxsize=2.9in\epsfbox{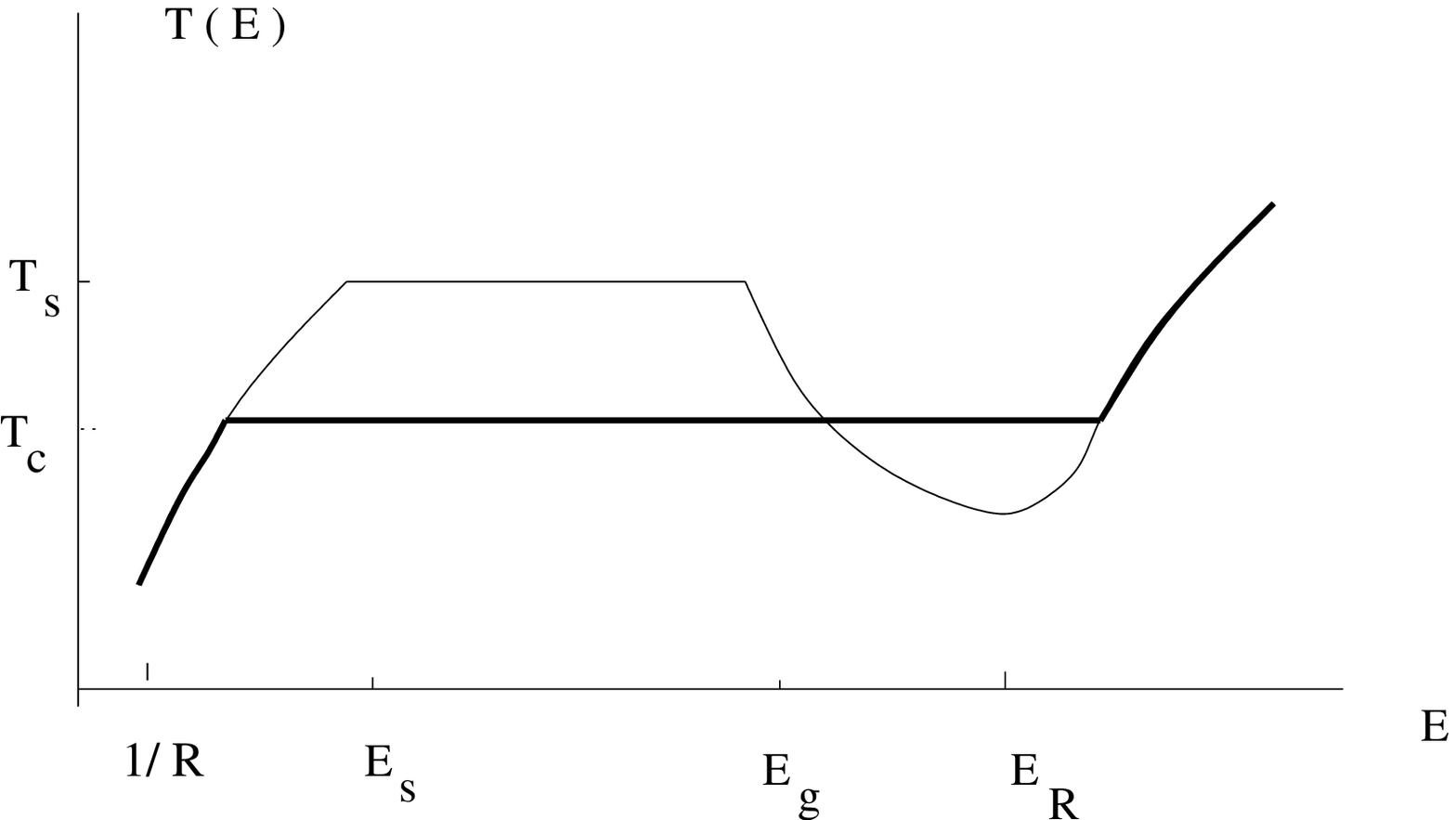}}
\caption{\sl The complete microcanonical temperature function for thermal
AdS spaces, in the four-component approximation. In addition to the
gravitons, heavy strings and small black holes, we now include large AdS
black holes. When small Schwarzschild black holes grow to size $R$, at
energies $E_{R} \sim N^2 /R $, their specific heat becomes positive and
the temperature can grow without bound with $T\sim E^{1 / d}$,
corresponding to the dual CFT in $d$ dimensions.
 A Maxwell construction (in the thick line) shows that the Hagedorn
plateau is only accessible to superheated states. A first-order  phase
transition at $T_c \sim 1/R$ nucleates very large black holes of mass
$M \sim N^2 R^3 T_c^4 $,  directly
out of the massless graviton phase.
 }
\end{figure}

\begin{figure}[ht]
\centerline{\epsfxsize=2.9in\epsfbox{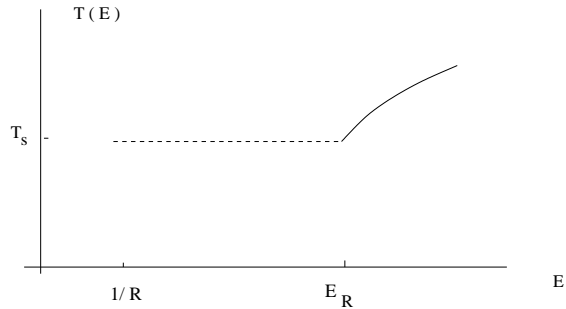}}
\caption{\sl At very weak coupling $g_s \sim 1/N$, the system matches the
perturbative regime of Yang--Mills theory and several hierarchical windows 
of the system
shut down. The ten-dimensional description
becomes strongly coupled and standard geometrical intuition breaks down. The
string energy $E_s$ that signals the beginning of the 
Hagedorn plateau becomes of the same
order as the finite-size gap of the gauge theory $E_{\rm gap} \sim 1/R$, so that
the phase of ten-dimensional graviton entropy disappears. 
The plateau ends at $E_R \sim N^2 /R$, the energy of the phase  transition   
into the Yang--Mills plasma. This threshold coincides with $E_g$, and the phase
of ten-dimensional black holes with negative specific heat also disappears. Instead,
the details of the plateau must be worked out in Yang--Mills  perturbation theory in the
't Hooft coupling $g_s N < 1$, where some ``precursors" of the strong-coupling
behaviour described in Fig 4 can be identified (c.f. \cite{aha}). 
  }
  \end{figure}

\begin{figure}[ht]
\centerline{\epsfxsize=1.9in\epsfbox{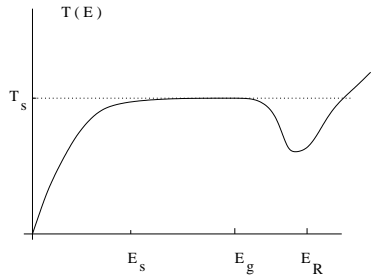}}
\caption{\sl The fine structure of a weakly metastable  Hagedorn plateau.
The Hagedorn temperature $T_s$ is strictly limiting in this region and the specific heat
on the approximate plateau is large and positive, becoming locally unstable
for $E>E_g$. Finite-size effects
imply this type of behavior for an ideal gas of closed strings.
}
\end{figure}

\begin{figure}[ht]
\centerline{\epsfxsize=1.9in\epsfbox{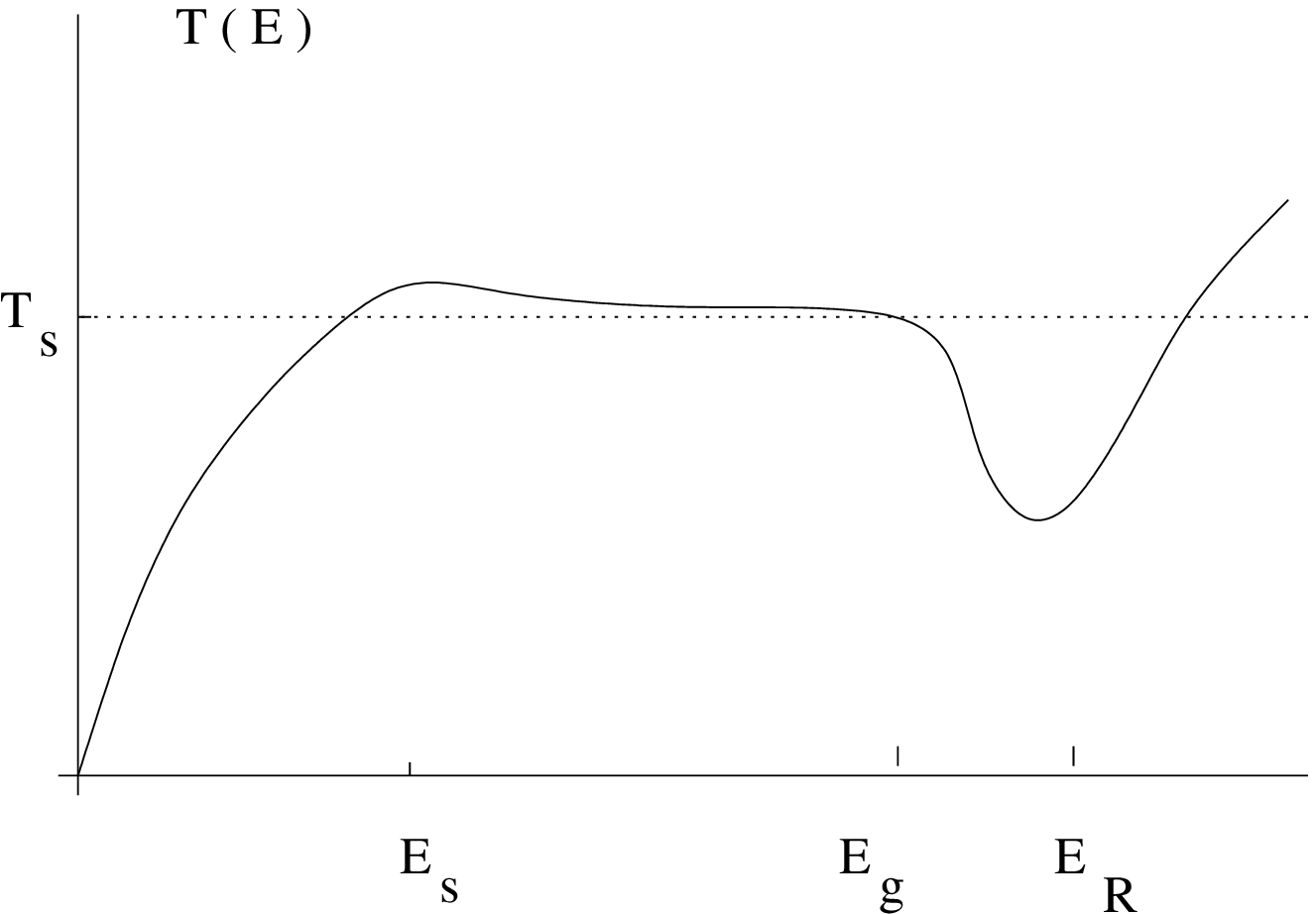}}
\caption{\sl The fine structure of a locally
unstable Hagedorn plateau. In this case
$T_s$ is first crossed at $E\sim E_s$ and then approached from above
 in the regime $E_s \ll E\ll E_g$. Local
instability sets in at $E\sim E_s$. The
 Hagedorn temperature is
still approximately maximal in the plateau region.
 This behavior appears  in the ideal gas
approximation for sufficiently large dimensionality, provided
finite-size effects can be neglected.
}
\end{figure}

In the larger picture that we are describing,
 the details of the Hagedorn plateau  at weak coupling  are not
very important. The Hagedorn temperature is approximately maximal, and only
accessible by means of  superheated states. However, in situations where
the string coupling can be taken very small at fixed $\ell_s$,
 the Hagedorn transient can
be enlarged, and it is interesting to work out the finer details of the hot
string gas. In figures 6 and 7 we depict the two main qualitative scenarios, corresponding
to a weakly metastable Hagedorn phase and a locally unstable Hagedorn phase,
respectively.  It turns out that
 boundary conditions  and finite-size effects start playing an
important role even in the ideal gas approximation. This is not very surprising,
since highly excited strings are ``long", macroscopic states, that are
sensitive to the large-scale structure of spacetime. This form of
UV/IR connection is the source of many subtleties in string thermodynamics. Some
of them will be described in the following section.

\section{Ideal gas of long strings}

\noindent

In this section we review, in rather picturesque manner,
 some  well-known  features  of string thermodynamics
in the free approximation. We start by introducing an intuitive geometrical picture
for a highly excited string as a random walk in target space. The large entropy factor
corresponding to the shape of the random walk in space explains  
why highly energetic strings dominate the thermodynamics in spite of their
large energy.   
In addition, the random walk picture becomes  very convenient to calculate and interpret
 the leading
corrections to (\ref{hagen}). In the approximation of free strings, these
corrections are determined by finite-size effects.   

\subsection{Random Walks}

\noindent

Consider a highly excited closed string represented as a random walk in target space.
 The energy $\varepsilon$ of
the string is proportional to the length of the random walk. The number of those
with a fixed starting point thus grows as $\exp(\beta_s \,\ve)$, explaining the bulk of
the entropy of highly energetic strings. Since the walk
must close on itself, this overcounts by a factor of roughly the volume of the
walk, denoted $V_{\rm walk} = W$. Finally, there is a factor of $V=L^d$ from the
global translation of the walk in a volume $V$,
 and a factor of $1/\ve$ because any point in
the string can be a starting point. The final result is 
\begin{equation}
\label{cld}
\omega_{cl} (\ve) \sim V\cdot {1\over \ve}\cdot {e^{\beta_s \ve} \over W(\ve)}
\;.
\end{equation}
There are two characteristic limiting cases. The volume of the walk is
of order $\ve^{d/2}$ when it is well-contained in  $d$ spatial dimensions (this corresponds
to $L\gg \sqrt{\ve}$), whereas
it saturates at order $V$ when it is space-filling $(L\ll \sqrt{\ve})$.
Hence, we find a density of states per unit volume 
\begin{equation}\label{clnc}
\omega_{cl} (\ve)/V \sim {e^{\beta_s \ve} \over \ve^{1+d/2}}
\end{equation}
in $d$ non-compact dimensions, and
\begin{equation}\label{ecs}
\omega_{cl} (\ve) = {e^{\beta_s \ve} \over \ve}
\end{equation}
in a compact space that contains completely the highly-excited string states.
In this case the formula (\ref{ecs}) gives the exact leading term of
the density of states, including the proportionality constant.     
  
The random walk picture is very geometrical and general. For example, it shows
that these densities are largely independent of spacetime topology 
 and only depend on the degree of ``containment"
of the random walk on the available volume.

 As an example of its generality,
we can also derive the corresponding entropies for highly excited open strings
in  a standard D$p$--D$q$ sector. In this case, 
we have the same leading exponential
degeneracy for random walks with a fixed starting point on the D$p$-brane. Fixing
the endpoint of a particular point of the D$q$-brane divides by a factor
of the total random walk volume $W$. Since endpoints can move in the part
of each brane occupied by the walk, we have a further degeneracy factor
$$
(W_{\rm NN} W_{\rm ND})\cdot (W_{\rm NN} W_{\rm DN})\,,
$$
where N and D refer to Neumann and Dirichlet boundary conditions. Finally,
the overall translation of the walk in the excluded NN volume gives a factor
$V_{\rm NN}/W_{\rm NN}$. The final result is
\begin{equation}\label{las}
\omega_{op} (\ve) \sim {V_{\rm NN} \over W_{\rm NN}} \cdot W_{{\rm NN}+{\rm ND}}
\cdot W_{{\rm NN} + {\rm DN}} \cdot {e^{\beta_s \ve} \over W} \sim {V_{\rm NN} \over
W_{\rm DD}} \,e^{\beta_s \ve}
\;.
\end{equation}
We see that the density of states is only sensitive to the volume of the
random walk in  the DD directions.  Again, we have two qualitatively limiting
cases: if the random walk is well-contained in the 
$d_\perp$ directions with  DD boundary conditions  we have
$L_{\rm DD} =L_\perp \gg \sqrt{\ve}$ and $W_{\rm DD} \sim \ve^{d_\perp /2}$ so that
\begin{equation}\label{opend}
\omega_{op} (\ve)/V_{\rm NN} \sim {e^{\beta_s \ve} \over \ve^{d_\perp /2}}
\;.
\end{equation}
On the other hand, if the walk is filling the DD volume we find
\begin{equation}\label{openc}
\omega_{op} (\ve)/{V_{\rm NN}} \sim {e^{\beta_s \ve} \over V_{\rm DD}} 
\;.
\end{equation}

These densities can also be obtained as equilibrium distributions
that solve Boltzman equations for interacting random walks (c.f. \cite{love}).
We may summarize the results  by the parametrization 
\begin{equation}\label{param}
\omega(\ve) \sim f \cdot {e^{\beta_s \ve} \over \ve^{1+
\gamma}}
\;,
\end{equation}
where $f = V_\parallel / V_\perp$. Here  
$V_{\parallel}$ is the volume available to the center of mass motion of
the walk and $V_\perp$ is the transverse volume in DD directions (we set $
V_\perp =1$ in string units when the DD directions are noncompact, or we have
closed strings). The  exponent in (\ref{param})  is $\gamma=-1$ for space-filling open
random walks and $\gamma =0$ for space-filling closed random walks. For 
open random walks that are well-contained in $d_\perp$ directions 
 we have $\gamma = -1 + d_\perp /2$.  

These finite-size effects  induce a negative logarithmic  correction
to the entropy of a highly excited string, which in turn gives it a negative
specific heat. Hence, long strings have a tendency to break the equipartition
of energy, which would flow into   
  one single long string. This also breaks extensivity. Consider two uncoupled subvolumes of the gas, 
each with its own single dominating long string.  Once brought together
there can be only one longest string violating extensivity.

\subsection{The full string gas}

\noindent

In
normal systems,
 the  thermodynamic or infinite-volume limit is a useful formal tool
in the study of the  phase structure. In the case of 
strings, their extended nature  puts into question usual assumptions about
extensivity of the thermodynamic functions, and  
the whole issue of the thermodynamic limit must be re-examined by
working at finite volume from the outset. However,
some  heuristic
rules of thumb can be envisaged without calculation.

 Consider an ensemble of long
closed strings with total available energy $E$. If all this energy flows
to a single long string, the random walk acquires
 size $E^{1/2}$ in string units.
The condition for the walk to be well-contained is thus $E^{1/2} \leq L$. 
A thermodynamic limit of large $L$
 with constant energy density $ \rho= E/ L^d$ is consistent with this condition only when 
$
1< \rho \leq L^{2-d}
$, 
which requires $d\leq 2$ 
(the first inequality follows from the condition of long-string dominance
over the massless modes). Hence, we obtain $d=2$ as the critical (spatial) 
dimension separating string gases with ``normal" thermodynamical behavior
from those that have important finite-size effects.

A convenient formalism to discuss the transition from the single long string
to a  gas of long strings is to obtain the full density of states $\Omega(E)$
from a formal partition function:
\begin{equation}\label{fomp}
Z(\beta) = \Tr \,\exp(-\beta\, H_{\rm SFT}) \equiv
 \int_0^\infty dE\,\Omega(E)
\,e^{-\beta E} \;,
\end{equation}
where $ H_{\rm SFT}$ denotes the second-quantized Hamiltonian of the full
string field theory and the trace is over the physical Hilbert space of
the full string theory. To leading order in perturbation theory,
 $ H_{\rm SFT}$ 
is the direct sum of the free-field Hamiltonians for each particle degree
of freedom of the single-string Fock space. The inverse ``temperature" $
\beta$ defines a consistent canonical ensemble only for $\beta >\beta_s$. 
Indeed, above the Hagedorn temperature, single string states  corresponding
to long random walks cause (\ref{fomp}) to diverge. Still, if $Z(\beta)$ is
defined by analytic continuation in the complex $\beta$ plane, we can write
an integral formula for $\Omega(E)$ as the inverse Laplace transform
\begin{equation}\label{invl}
\Omega(E) = \int_{C_\beta} {d\beta \over 2\pi i} \;e^{\beta\,E} \,Z(\beta)
\;,
\end{equation}    
where the contour $C_\beta$ is parallel to the imaginary axis and to the
right of all singularities of $Z(\beta)$.  The entropy of the string
gas is $S(E) = \log\,\Omega(E)$ and is determined by the singularities of
$Z(\beta)$, upon evaluation of the integral (\ref{invl})
 by contour deformation \cite{deo, usabel}.

Although $Z(\beta)$ can be evaluated explicitly in the one-loop approximation \cite{polc}, we
follow here a more heuristic route. In estimating  $Z(\beta)$ near the Hagedorn singularity,
  we can assume Maxwell--Boltzman statistics, because the
dominating long strings are  macroscopic, and thus they behave quasiclassially. We  may
 then  write
$Z(\beta)= \exp\,z(\beta)$, where $z(\beta)$ is the single-string partition
function. It is related to the single-string density of states by the
Laplace transform  
\begin{equation}\label{stpt}
z(\beta) = \int_0^\infty d\ve\,\omega(\ve) \,e^{-\beta\,\ve}
\;.
\end{equation}
By direct calculation, we find that the behavior of $z(\beta)$ near the
Hagedorn singularity $\beta=\beta_s$ is given by  
\begin{equation}\label{crit}
z(\beta) \sim f\,(\beta-\beta_s)^{\gamma} \,\left[\log(\beta-\beta_s)
\right]^\delta
\;,
\end{equation}
where $\delta=1$ if $\gamma$ is a non-negative integer
 and $\delta=0$ otherwise. We see that
the   Hagedorn densities (\ref{param}) are associated to critical
behavior as a function of the formal canonical temperature $1/\beta$, with
a critical exponent given by  $\gamma$, as in (\ref{param}).   

One can evaluate the integral (\ref{invl}) in various approximations, depending
on the different dynamical regimes of energy and  volumes. Whenever
the saddle-point approximation is applicable, one finds an equivalence between
canonical and microcanonical ensembles, with positive and large
 specific heat. A
necessary condition for this is that $\gamma \leq 1$, ensuring that the
canonical internal energy 
$
E(\beta) \sim \pt_\beta \,z(\beta)
$ diverges at the Hagedorn singularity. Looking at the values of $\gamma$
as a function of the dimensions, we see that stable canonical behavior
is to be obtained for closed strings in low-dimensional thermodynamic limits,
 $d\leq 2$, or open strings with $d_\perp \leq 4$ noncompact DD  
dimensions, i.e. D$p$-branes with $p\geq 5$ and noncompact transverse 
dimensions.  In all these cases, the energy satisfies the usual
laws of equipartition in terms of the individual strings.  

     In the cases that a  
saddle-point approximation is not available, one can either evaluate
the integral exactly in special marginal 
cases (in particular for $\gamma=0$),
 or find a complementary approximation (see \cite{deo, usabel} for a summary of
 cases). For example, in certain situations one can perform 
an expansion in powers of the single-string partition function, 
$$
Z(\beta)
\approx Z(\beta_s) \,[1+ z(\beta)-z(\beta_s)] +  O\left(|z(\beta)-z(\beta_s)|^2 \right)
\;.$$
 In this case, one
finds single-string dominance with negative specific heat. The paramount
example of this behavior is that of D$p$-branes with $p<5$ in ten non-compact
dimensions. 
Naively, closed strings with $d>2$ also belong to this category. However
we noticed before that the thermodynamic limit of closed strings is sensitive
to finite-size effects precisely in this regime of dimensions, so that
the large-volume limit must be studied with due attention to these 
boundary effects. Similarly, open random walks with energy $E> L_\perp^2$ 
are also subject to  finite-size effects. 

This situation can be illustrated by considering the energy distribution of
single long strings in the gas \cite{deo, usabel}. The number of states with a long    
 string  of energy  $\ve$,
 at fixed total
energy $E$,  is proportional
to
\begin{equation}\label{stes}
D(\ve, E) = {\omega(\ve) \Omega(E-\ve) \over \Omega(E)}
\,.
\end{equation}
In systems where the finite-size effects can be neglected there is a tendency
for the energy to be carried   
dominantly  by a single long string, as in figure 8, whereas the energy
is uniformly distributed  when the long strings are constrained by the available
volume, c.f. figure 9.  

\begin{figure}[ht]
\centerline{\epsfxsize=2.9in\epsfbox{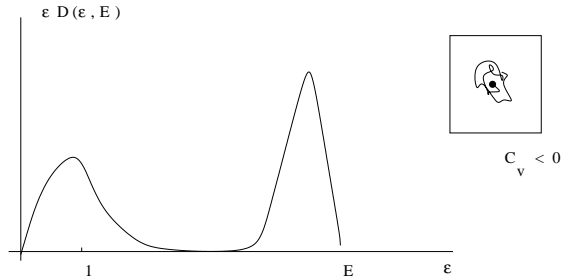}}
\caption{\sl Single-string  energy distribution  for a total energy E,
in systems where the random walks are well-contained and with
 large co-dimension. 
The first peak corresponds to the energy in massless modes. The area below it
 represents the
energy deposited in these modes, of order 
$\rho_s V$. The second peak corresponds to a single long string that captures $E-\rho_s V$
of the energy. This situation corresponds to an unstable Hagedorn phase, as in figure 7.   
 }
\end{figure}

\begin{figure}[ht]
\centerline{\epsfxsize=2.9in\epsfbox{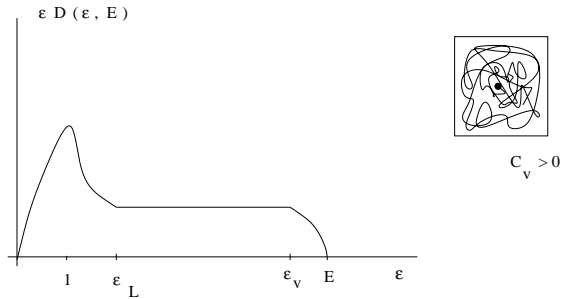}}
\caption{\sl Single-string energy distribution for systems with a 
 very ``dense packing" of random walks
in the allowed volume. The initial peak of the massless modes decays to a plateau that 
starts at $\ve = \ve_L \sim L^2$, the energy of volume-saturating random walks. The
plateau continues up to energies $\ve_v \sim E-\rho_s V$. This situation corresponds to
a locally stable Hagedorn phase, as in figure 6.   
 }
\end{figure}

Systems with close-packing of  random walks (high energy in
a fixed volume) have $\gamma=-1$ for
open strings and $\gamma=0$ for closed strings. In the first case,
the saddle point approximation applies and we have a gas of open strings
with canonical behavior, positive specific heat and entropy of the form
$$
\Omega (E)_{\rm open} \sim \exp \left(\beta_s E + C \sqrt{E} \right)
$$
with  some constant $C$. The case of closed strings is slightly more involved. 
The leading singularity at very high energy and finite volume $V=L^d$
is always a simple pole of the partition function at the Hagedorn singularity,
\begin{equation}\label{singlp}
Z(\beta) = (\beta-\beta_s)^{-1} \cdot  Z(\beta)_{\rm regular} 
\;.
\end{equation}
This pole alone produces a multistring density
\begin{equation}\label{mmi}
\Omega(E)_{\rm closed} \sim \exp\left(\beta_s E + \rho_s V  \right) 
\;,
\end{equation}
with $\rho_s = O(1)$ in string units.  Hence, the specific heat is still
infinite in this approximation. The contribution of the subleading
singularities turns the thermodynamics into a weakly limiting behavior
with positive specific heat and exponentially suppressed corrections to
the linear entropy law. This conclusion can be anticipated by the
study of the energy distribution of single long strings. Calculating 
the distribution function $D(\ve, E)$ from (\ref{ecs}) and (\ref{mmi}) one finds
$D(\ve, E) \sim 1/\ve$, so that the energy distribution $\ve \,D(\ve, E)$ is flat
for $\ve > L^2$, suggesting equipartition and positive specific heat.   
 
It is interesting to notice that, on volumes of the order of the
 string scale, the 
 entropy in open strings  grows as 
$$
S(E)_{\rm open} \sim \beta_s E + C \,\sqrt{E}\,,
$$
 whereas closed strings are marginally limiting, with  
$$
S(E)_{\rm closed} \sim \beta_s E - C' \, E^{16} \,e^{-\eta E}
\,,$$
where $C, C'$ and $\eta$ are $O(1)$ constants in string units. This means that 
primordial cosmology scenarios   which start
with a ``small" universe (c.f. \cite{branvafa}),
 are very sensitive at the possible presence of D-branes in the
primordial ingredients, since open strings
 dominate the density of states in these circumstances \cite{usabel, obv}.

\section{The thermal scalar}

\noindent

The critical behavior apparent in equation (\ref{crit}) 
 begs for a representation in terms of the dynamics of
light modes but, what light modes could possibly have a bearing on this
 situation,
since we are looking at extremely massive string states from the beginning?

An interesting answer can be obtained by a formal detour.
We first notice that the random walk in spatial dimensions is the same as the
configuration space of a path integral for a relativistic particle in Euclidean
space. In this picture, 
   $\ve$ is    proportional to the length of the walk, and (\ref{stpt}) is a
Schwinger representation of a random world-line of length $\ve$. 

Let us consider first the case of a closed random walk and write (\ref{crit}) 
using (\ref{stpt}) and (\ref{cld}), 
\begin{equation}\label{forml}
\log\,Z(\beta) \approx z(\beta) 
 \sim V\,\int_0^\infty {d\ve \over \ve}\, {e^{-(\beta-\beta_s)\ve} 
\over W(\ve)} \;.
\end{equation} 
For the case of a well-contained walk we have 
$$
W(\ve)^{-1} \sim \ve^{-d/2}\sim  \int {d^d k \over (2\pi)^d} \;e^{-\ve k^2}
$$
so that the complete expression (\ref{forml})
 is just a Schwinger proper-time representation
of a one-loop determinant for a scalar field in $d$ dimensions \cite{brab}, 
\begin{equation}\label{dett}
\log\,Z(\beta) = -{n_{\rm eff} \over 2} \,\Tr \,\log \,\left[-\nabla^2
 + m^2_{\rm eff}(\beta)\right]_{{\bf R}^d} 
\;,
\end{equation}
where the effective mass $m^2_{\rm eff} \sim (\beta-\beta_s)$ is indeed vanishing
at the Hagedorn temperature. We have parametrized the normalization of the
partition function by the number $n_{\rm eff}$, an effective number of 
components of the thermal scalar. 

We see that the critical behavior of very long strings
can be {\it formally} parametrized by the dynamics of a light scalar in $d$
Euclidean dimensions. In the ideal-gas   approximation considered here,
the random walk is identified with the Feynman path of the light scalar particle
in Euclidean space.  

\begin{figure}[ht]
\centerline{\epsfxsize=2.2in\epsfbox{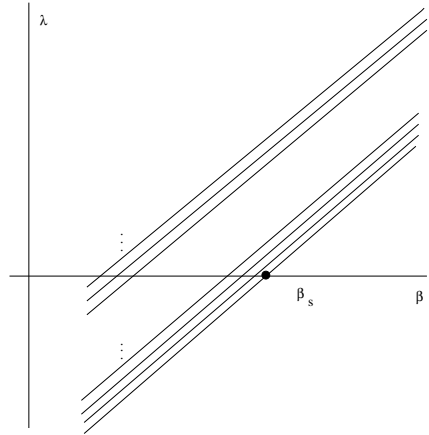}}
\caption{ \sl The eigenvalue spectrum of the operator $-\nabla^2 + m_{\rm eff}^2 (\beta)$ in
the vicinity of $\beta \approx \beta_s$. As $\beta$ decreases, there is a lowest eigenvalue
that vanishes at $\beta=\beta_s$. For a finite box of unit size in string units, the eigenvalue
spacing is of $O(1)$ and subleading singularities appear below $\beta_s$ 
separated by
$O(1)$ intervals.  For a finite box of size $L\gg \ell_s$, there are  associated bands of momentum
modes with spacing of order $1/L^2$. In the limit $L \rightarrow \infty$ the bands become
quasicontinuous.  
 }
\end{figure}

 For random walks in finite volume, the  spectrum of the operator
$-\nabla^2$ 
is gapped with characteristic scale $1/L^2$, so that the momentum integrals
are irrelevant for  $\beta-\beta_s \ll 1/L^2$, leading to a purely
 logarithmic free
energy for $\beta$ sufficiently close to $\beta_s$.   
  In this case  the basic canonical singularity for  
closed strings in finite volume takes the form
\begin{equation}\label{bsin}
Z(\beta)_{\rm sing} \sim {1\over (\beta-\beta_s)^{n_{\rm eff}/2}} 
\;. 
\end{equation}
Agreement with (\ref{singlp}) requires $n_{\rm eff} =2$, i.e. the thermal
scalar can be considered as a {\it complex} field. An explicit stringy
construction in the next section will confirm this conclusion. 
 The opening
of $d$ dimensions in the limit $L\rightarrow \infty$ corresponds to a
dense set of  poles separated by a distance of $O(1/L^2)$ in $\beta$-space,
accumulating at $\beta=\beta_s$ and transforming the pole singularity at $
\beta_s$ into
a cut of the form (\ref{crit}).    

\begin{figure}[ht]
\centerline{\epsfxsize=2.9in\epsfbox{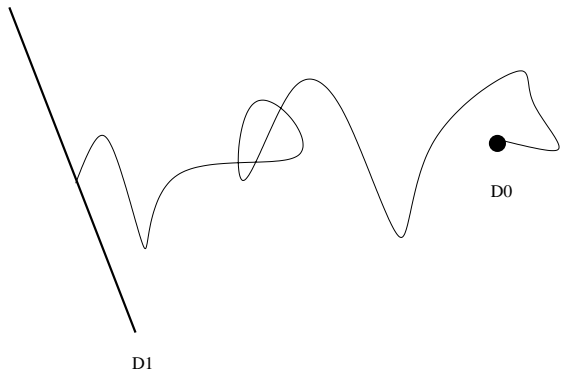}}
\caption{\sl The random configurations of a long string in {\it space} are identified with the
random  paths of of an Euclidean field $\chi$. Here we show a long open string between
D1 and D0 branes or, equivalently, a contribution to the propagator of the thermal scalar.    
 }
\end{figure}

The effective thermal scalar description also explains the critical behavior of open strings
on D-branes. The geometry of the random walk suggests that the relevant
quantity is now the propagator between boundary states:
\begin{equation}\label{pro} 
\log Z(\beta)_{pq} \sim \left\langle {\rm D}q\,\Big| \,{1\over -\nabla^2 + m^2_{\rm
eff}(\beta) }\,\Big|\,{\rm D}p\,\right\rangle\;,
\end{equation}
where D-brane boundary states project onto the zero-momentum sector
in NN directions, and absorb any momentum in the DD directions. Hence, the
momenta flowing through the propagator are only those in the DD directions, 
 explaining the factor of  $\ve^{-d_\perp /2}$ in the number of states' densities.  

\subsection{Stringy origin of the thermal scalar}

\noindent

The thermal scalar formalism operates in the {\it spatial} Euclidean space.
However, being formally a path-integral representation of a canonical
partition function, note that the thermal circle ${\bf S}^1_\beta$
of period $\beta$ does not feature explicitly in the formalism.   
In fact, it turns out that the thermal scalar is an effective description
 in which
the thermal circle  has been integrated out,
 and the thermal scalar is
one of the light
 ``degrees of freedom" that remains when $\beta \approx \beta_s$.  
As we shall see now, one can ``integrate in" the information regarding
${\bf S}^1_\beta$, and the result is rather surprising. 

To argue this point in the simplest example,
 consider the one-loop free energy of bosonic open strings. The generalization
 to
superstrings is straightforward and brings no new conceptual issues.  At this
level we can consider such a free energy as the
 sum of free energies for each physical particle
in the open-string spectrum, so we have an expression of the form
\begin{equation}
\log\,Z(\beta)_{op} =
 -\shalf \,\Tr_{op} \,\log\,\left[ -\pt^2 + M^2 \right]_{X_\beta} 
\end{equation}
where the trace sums the spectrum of the operator $-\pt^2 +M^2$, over all
the open-string fields. The kinetic operator $-\pt^2$ is defined
on the thermal manifold $X_\beta = {\bf S}^1_\beta \times {\bf R}^d $ as 
$$
-\pt^2 = -\pt_\tau^2 -\nabla^2
\;,$$
where $\tau$ is a coordinate on ${\bf S}^1_\beta$ and $-\nabla^2$ is
the standard Laplacian on ${\bf R}^d$.   
  Passing to a Schwinger proper-time representation
and isolating the discrete eigenvalues of  $-\pt^2_\tau =  4\pi^2 n^2 /\beta^2$, 
  we get contributions  of the form 
\begin{equation}
\log\,Z(\beta)_{op} = \shalf \,
\int_0^\infty {dt \over t} \left[\cdots \sum_n e^{-2\pi t(4\pi^2 n^2/\beta^2 +
\dots)} \right]   
\;.\end{equation}
In this formula and the rest of the section, we pay no attention to the
ultraviolet divergences at small Schwinger parameter,
 since the $\beta$-dependent part of the partition function is finite.
To this end,  we adopt the prescription of
subtracting the contribution of the vacuum energy from $\log\,Z(\beta)$.

In a world-sheet 
path-integral picture, 
we have a one-loop vacuum diagram of open strings, the annulus with modular
parameter $t$. The momentum
modes in the thermal circle can be transformed into winding modes of the string world-sheet around
the thermal circle by   
Poisson resummation in the index $n$, resulting in terms of the form
\begin{equation}\label{insw}
{1\over \sqrt{t}} \sum_\ell e^{-\beta^2 \ell^2 / 8\pi t}
\;.
\end{equation}
We can see that $\ell$ is a winding number by constructing the embeddings of the annulus that
wrap $\ell$ times on the thermal circle,
$
\tau  = {\ell \beta \sigma_2 /  t} 
$, 
with action
$$
S_\ell = {1\over 4\pi} \int_0^\pi d\sigma_1 \,\int_0^{2\pi t} d\sigma_2 \,\left(
 {\pt \tau \over \pt\sigma_2} \right)^2 =
{\beta^2 \ell^2 \over 8\pi t}
,$$
so that we recover (\ref{insw}) as a semiclassical sum 
 $\sum_\ell \exp(-S_\ell)$.   
 A further 
 change of variables $s= 1/t$,
 represents the modular transformation to the closed-string channel,
in which we see a tree-level cylinder diagram of closed strings with
modular parameter $s$. The closed strings carry now winding
modes around the thermal circle. 
Detailed inspection of these manipulations in the complete expression above shows that all powers of
$s$ can be exponentiated in the proper time integral so that one finds 
\cite{mav} 
\begin{equation}
\log\,Z(\beta)_{op} \sim \int_0^\infty ds \,\Tr_{cl} \,
\sum_\ell \exp\left[-{\pi\over 2}\,s\, \left(-\nabla^2 + M_\ell^2 (\beta) \right)
\right]
\;,\end{equation}
which has the form of a proper-time representation of a propagator, 
just like (\ref{pro}). The trace $\Tr_{cl}$ runs now over the whole tower
of closed-string
states.  The
crucial point is the emergence of effective mass terms proportional to
 $\beta^2$, i.e. we have 
$$
M^2_\ell = {\beta^2 \ell^2 \over 4\pi^2} + \dots\ 
,$$
  which can be interpreted as  
winding modes of closed strings on ${\bf S}^1_\beta$. The lightest of these winding modes, corresponding
to $\ell = \pm 1$, can be assembled into a complex scalar field $\chi$. Now, since the thermal
circle breaks supersymmetry, the spectrum of the operator $M_\ell^2$ in these winding sectors is not    
guaranteed to be positive definite, but has actually a negative lowest eigenvalue $-|M_0|^2$. Defining
$\beta_s = 2\pi |M_0|$ we have an effective mass
$$
m^2_{\rm eff} (\beta) = {\beta^2 - \beta_s^2 \over 4\pi^2}
\;,$$
as required for the thermal scalar. Hence, we see that the stringy origin of the thermal scalar
is in the thermal winding modes of the closed-string sector. The critical behavior arises because
this thermal scalar becomes massless at $\beta = \beta_s$.  

\subsection{Is the thermal scalar ``physical"?}

\noindent

The parametrization of long-string critical behavior in terms of an effective
thermal scalar poses the question of the
 physical interpretation of these degrees
of freedom. In the path-integral picture they arise naturally as closed-string
winding modes around the thermal circle. However, the Hamiltonian
interpretation of these modes is not immediate,
 and therefore its physical status
remains somewhat unclear. 

In open-string sectors modular covariance (open-closed string duality)
 provides an answer to the previous question. 
The closed-string winding modes in the closed-string channel are  dual under Poisson
resummation of the 
ordinary  momentum modes  of open strings in the crossed channel. Hence, the physical
Hamiltonian only has open-string modes and no states of the field $\chi$ are visible
among the physical open-string spectrum. 

On the other hand, the answer is less obvious in the closed-string sector.
 The one-loop
diagram of closed strings has two independent winding modes, one for each of the
cycles of the worldsheet torus. One set of winding modes can be interpreted as a
Poisson dual of standard momentum modes on ${\bf S}^1_\beta$. However, another set
of winding modes remains and we still need to find a  Hamiltonian 
interpretation for those. 
In such a Hamiltonian interpretation, they would appear as ``timelike"
winding modes, a rather mysterious notion. 

The resolution of this paradox is again related 
to modular invariance. Following the ``empirical" 
 reasoning of the previous section, 
we start from the physical Hamiltonian picture, i.e. we
consider the  thermal free energy in the ideal gas approximation as
 given by the
sum of free energies for all field degrees of freedom in the spectrum:
\begin{equation}\label{analogo}
\log\,Z(\beta) = \log\,\Tr_{\rm SFT} \,e^{-\beta \,H_{\rm SFT}} =
- \sum_{f} \Tr_{(f)}  \,\log\,\left(1-
e^{-\beta \,\omega_f}\right) 
\,,\end{equation}
where $\Tr_{(f)}$ runs over the momentum
 and spin degrees of freedom of each field in the
string spectrum. 
 Standard manipulations yield an expression in terms of determinants on  
the Euclidean manifold $X_\beta ={\bf S}^1_\beta \times {\bf R}^d$,   
\begin{equation}\label{det}
\log\, Z(\beta) =- \shalf \sum_f \Tr \,
\log\,\left[-\pt^2 + M_f^2 \right]_{X_\beta 
} 
\,,\end{equation}
where now the trace runs over the spectrum of the kinetic operator on the Euclidean manifold
$X_\beta$. In a Schwinger representation,
\begin{equation}\label{scho}
\log\,Z(\beta) =  \shalf \int_0^\infty {d\tau_2 \over \tau_2} \,\sum_f \,\Tr 
\,\exp\left[-{\pi \over 2} \tau_2 \,(-\pt^2 + M_f^2 )_{X_\beta } \right] 
\,.\end{equation}
Now we notice that $\shalf (-\pt^2 + M^2) = L_0 + {\bar L}_0$,
 with $L_0$ the chiral
world-sheet Hamiltonian of the string theory. 
Introducing a level-matching constraint   
$$
\delta_{(L_0, {\bar L}_0)} = \int_{-1/2}^{1/2} d\tau_1 \,e^{2\pi i \tau_1\,
(L_0 - {\bar L}_0)} 
\,,$$
we can write a formal expression 
\begin{equation}\label{lo}
\log\,Z(\beta) =\shalf \int_S {d^2 \tau \over \tau_2^2} \;\tau_2\, 
\Tr_{\rm CFT} \;q^{\,L_0} \;{\bar q}^{\,{\bar L}_0}
\;,\end{equation}
where $q=e^{2\pi i \tau}$, with $\tau = \tau_1 + i \tau_2$. 
 Equation (\ref{lo})
 resembles the  string world-sheet partition function, except for the fact that
the integration domain $S: \tau > 0, \,-\shalf < \tau_1 
< \shalf$ does not coincide with the fundamental domain of the modular group
$F: |\tau| > 1, \,-\shalf < \tau_1 < \shalf$. In addition,
 the trace only includes {\it physical} modes on ${\bf R}^d$ that were
already traced over in the statistical sums (\ref{det}) and
(\ref{scho}), i.e.,  
there are  momentum modes on ${\bf S}^1_\beta$, but no winding modes. In
particular, this implies that the integrand cannot be modular invariant.

It turns out that these two facts essentially cancel one another. One can
trade the summation over thermal winding modes by an extension of the
integration region from the fundamental domain $F$ to the strip $S$ (c.f.
\cite{roth}). The basic identity is
\begin{equation}\label{mro} 
\sum_{(\ell, \ell')
\neq (0,0)} \int_F {d^2 \tau \over \tau_2^2} \,f(\tau,{\bar\tau}) \;
e^{-{\beta \over 4\pi\tau_2} \,|\ell' + \tau \ell|^2} =
\sum_{\ell \neq 0}
 \int_S {d^2 \tau_2 \over \tau_2^2} \,f(\tau,{\bar \tau}) \;e^{-{\beta \over
4\pi \tau_2} \,\ell^2}
\;,
\end{equation}
where $f(\tau, {\bar \tau})$ is modular invariant. The restriction in the
winding sums in (\ref{mro}) ensures that the vacuum energy,
or $\beta\rightarrow \infty$ limit, 
is subtracted, a necessary condition  for the theorem to hold. 
 Equivalently, we can  
use the normal-ordered $H_{\rm SFT}$ in (\ref{analogo}), although 
in the case of superstrings
this term  vanishes from the outset. 
 Thus, modular invariant expressions always have
two sets of thermal winding modes, but  ``Hamiltonian" representations
automatically unwrap one of these sets. The remaining one can be dualized
into standard momentum modes on ${\bf S}^1_\beta$, as in the previous
subsection.      

There are some subtleties that we have chosen to hide.  
For example, the identity (\ref{mro}) must be used for  
  $\beta > \beta_s$, due to convergence
problems that can affect
 the analytic structure of $\log Z(\beta)$ \cite{mavoso}.

The conclusion is that the thermal scalar does not have a strict physical
interpretation in terms of the particle degrees of freedom of the
original thermal gas. Rather, it is a formal
device which reconciles the subtleties of stringy modular invariance with
the rules of effective field theory.  

\section{Tachyon dynamics and the Hagedorn transition}

\noindent 

We have seen that the canonical formalism just below the
Hagedorn temperature, $0<\beta-\beta_s \ll 1$, involves an effective
light field whose quanta are topological winding modes that lack a 
direct Hamiltonian interpretation. 
We can write an effective action for
{\it static} and {\it spatial-dependent} configurations,
 in the spirit of Landau's mean-field formalism, 
\begin{equation}\label{eft}
S[\chi]_{\rm eff} = {\beta \over 2g_s^2} \int_{{\bf R}^d}
 \left( |\nabla \chi|^2 + 
 V_{\rm eff} (\chi^* \chi) + \dots \right)
\,,
\end{equation}
where
\begin{equation}\label{potta}
V_{\rm eff} (\chi^* \chi) = {\beta^2 - \beta_s^2 \over 4\pi^2} \,|\chi |^2 + {\rm interactions}
\end{equation}
and  the dots stand for
 contributions of other light degrees of freedom, such as
the ordinary massless modes of the string spectrum. 
A potential has been included to account for interaction corrections. Since the critical
behavior is
 characterized by $\chi$ becoming massless at $\beta = \beta_s$, the 
``post-Hagedorn" regime is formally described by a tachyonic $\chi$ field. In this case,
it is natural to associate the dynamics of ``tachyon condensation" with a phase
transition at the Hagedorn temperature. 
This ``dynamics" remains somewhat formal, since the interpretation of the effective
action (\ref{eft}) as a thermal effective potential is only valid for static configurations.
Therefore, the ``rolling down" along 
the tachyonic potential is not to be seen as a process
in real time. Rather, we just compare the free energies of static configurations with
different  values of $|\chi|^2$, averaged  over ${\bf R}^d$. These static backgrounds, being
off-shell, can be seen as building a renormalization-group flow on the string world-sheet.  

It turns out that the picture of a ``tachyon roll" down an unstable
potential is  separated from the perturbative phase $\bra |\chi|^2 \ket =0$   
by a first-order phase transition. The authors of \cite{aw} noticed
that the interaction of certain gravitational moduli with the thermal
scalar induce an unstable quartic term in the effective potential. The
full low-energy effective action on ${\bf R}^d$ 
 includes  the ten-dimensional dilaton
and graviton fields coming from the dimensional reduction on ${\bf S}^1$  
\begin{equation}\label{toteff}
S_{\rm eff} \sim {\beta\over g_s^2} \int_{{\bf R}^d}  
e^{-2\phi} \,\sqrt{|g|} \left[ -R-4(\pt \phi)^2 + |\pt \chi|^2 + 
{\beta^2 g_{\tau\tau} - \beta_s^2 \over 4\pi^2} \,|\chi|^2  + \dots \right]
\end{equation}
where the term $g_{\tau\tau} \beta^2$ represents the effect of fluctuations
in the proper length of the thermal circle (see figure 12).
 Since this is interpreted
as a local inverse temperature, (\ref{toteff}) incorporates the effect of
local temperature fluctuations in the mean-field approximation.

\begin{figure}[ht]
\centerline{\epsfxsize=1.9in\epsfbox{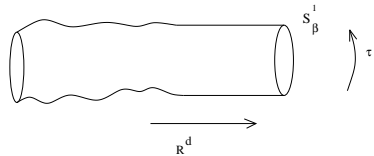}}
\caption{\sl The substitution $\beta^2 \rightarrow \beta^2 g_{\tau\tau}$ has the effect
of including local temperature fluctuations on $X_\beta$. 
}
\end{figure}

 Writing
$g_{\tau\tau} = 1+ \sigma$ and expanding the $\sigma$ dependence in a 
weak-field approximation, we have a tree-level coupling of the form
$\sigma \,\chi^* \,\chi$ which is not suppressed by derivatives or
by powers of $\beta-\beta_s$. This coupling is real, so that integrating
out the $\sigma$ field at tree level gives a negative-definite quartic
coupling for $\chi$ of the form
\begin{equation}\label{efaw}
V_{\rm eff} \sim -{\beta \over 2g_s^2} \left\langle \,\chi^* \chi \,\Big| 
\,{1 \over -\nabla^2} \,\Big| \,\chi^* \chi\,\right\rangle
\;.\end{equation}
This term is non-local because the field $\sigma$ is massless. It is
infrared divergent when evaluated on constant $\chi$ configurations.
However, a finite-volume regularization with a gap in the spectrum
of $\nabla^2$ renders it well-defined. The resulting picture is
represented in figure 13. The effective potential describes a
first-order phase transition occurring slightly below the Hagedorn temperature,
when the thermal scalar perturbative vacuum $\bra |\chi |^2 \ket =0
$ is still locally stable.
\vspace{-3mm}  
\begin{figure}[ht]
\centerline{\epsfxsize=2.7in\epsfbox{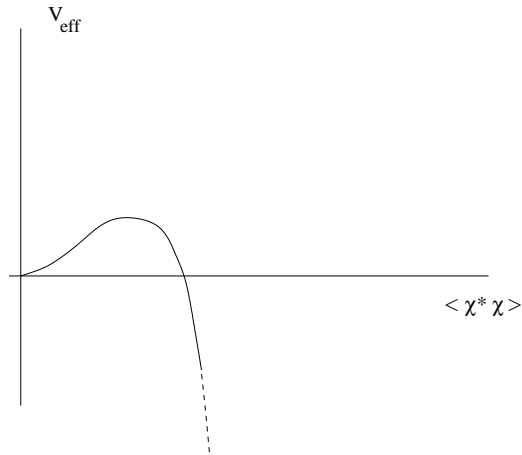}}
\caption{\sl The qualitative form of the thermal scalar
 potential for $\beta > \beta_s$. 
}
\end{figure}
 
\vspace{-6mm}  
\subsection{AdS regularization and Euclidean black holes}

The  Atick--Witten 
transition proceeds by nucleation of a domain with $\bra |\chi |^2 \ket
\neq 0$, followed by a ``roll" down the effective potential, a classical
process  that could perhaps be described in terms of world-sheet
 renormalization-group flows, in analogy with similar processes in the
decay of D-branes \cite{sen} and tachyonic orbifold singularities \cite{aps}. 

The particular  endpoint of the condensation process depends on the
mechanism of 
stabilization of the potential (\ref{efaw}). This should be related to
the details of the 
 finite-volume regularization, but we can expect in any case that 
the means to resolve 
this problem remains beyond the weak-field methods that lead to (\ref{efaw}).   
The related question  of what physical interpretation  should we assign to
the thermal scalar condensate $\bra |\chi|^2 \ket$ 
 becomes especially
acute in view of 
 the considerations of the previous
subsection on the ``formal" nature of the thermal winding modes.

In order to answer these questions, even at the heuristic level, we must
go back to the physical picture of section 1. The main lessons of the
microcanonical approach are the following:

{\it (i)} The naive infinite volume  thermodynamical
 limit is inconsistent with the instabilities of
gravity. Even the mild finite-size effects of free long strings   are important given
the marginal instability of the Hagedorn regime.  This means that the mean-field approximation
based on the thermal scalar effective action on $X_\beta = {\bf S}^1_\beta \times {\bf R}^d$
must be supplemented with an appropriate infrared cutoff.

{\it (ii)} Using AdS  spaces as an infrared regulator, we have  non-perturbative  physical
intuition in terms of the thermodynamics of the dual CFT, specified
at short distances by a gauge field theory. The Hagedorn gas of long strings
becomes a metastable superheated state.

{\it (iii)} The microcanonical picture of the first-order phase transition in AdS-regularized spaces
 involves black hole
nucleation, with latent heat of order $1/g_s^2$ (c.f. figure 4).

Let us consider then the AdS regularization of a ten-dimensional string gas in the standard 
 ${\rm AdS}_5 \times {\bf S}^5$ background of type IIB string theory. The metric takes the
form
\begin{equation}\label{adsm}
ds^2 = -\left(1+{r^2 \over R^2}\right) dt^2 + \left(1+{r^2 \over R^2} 
\right)^{-1}\;dr^2  + r^2 \,d\Omega_3^2 + R^2 \,d\Omega_5^2
\;,\end{equation} 
with $R^4 = g_s N$ in string units and $N$ the quantum of RR flux on ${\bf S}^5$. At a non-perturbative
level, this system is defined by the quantum mechanics of $SU(N)$ super Yang--Mills theory on a
three-sphere of radius $R$ and with Yang--Mills coupling $g^2 = g_s$. For $g_s N \gg 1$, the metric
is approximately flat on scales small compared to the radius of curvature $R$. A thermal ensemble defined
with respect to the time variable in (\ref{adsm}) has local temperature 
$$
T(r) = {T(0) \over \sqrt{1+ r^2 /R^2}}
\,.$$
Hence, on scales $\ell_s \ll r\ll R$   we have a macroscopic quasi-flat region with gas at temperature
$T\approx T(0)$. The gravitational redshift freezes the temperature on scales larger than the radius of curvature,
and we see that AdS works like a finite box of size $R$ as far as thermodynamics is concerned. The Euclidean
manifold describing this ensemble is $X_{\beta}$, obtained from (\ref{adsm}) by the standard Wick rotation
$t \rightarrow i\tau$, followed by the periodic identification $\tau \equiv \tau + \beta$, with 
$\beta = 1/T(0)$.
 This manifold looks locally like ${\bf S}^1_\beta \times {\bf R}^9$ on scales $r\ll R$. 
In particular, if $\beta \approx \beta_s$, we have a regularized version of the standard thermal manifold
with light thermal winding modes localized on the region $r< R$ (thermal winding modes supported at $r\gg R$ have
mass proportional to $\beta(r) = \beta \sqrt{1+r^2/R^2} \gg \beta_s$). 

The  reasoning of ref. \cite{aw} 
can be applied to type IIB strings on $X_\beta$ and one expects a first-order
phase transition towards a background with non-vanishing values of the thermal scalar $\bra |\chi|^2 \ket 
\neq 0$. Now, in order to understand the geometrical interpretation of this condensation process, we
can simply look at the endpoint of
 the decay in the dual CFT. This should be the thermal equilibrium state
at the corresponding temperature. Since we have $T\sim T_s \gg 1/R$ in the decay of the superheated ``Hagedorn"
states, the endpoint is the plasma phase of the CFT with entropy of $O(N^2) = O(g_s^{-2})$. In the gravity
 description, this is the large AdS black hole at temperature $T \gg 1/R$
\cite{adshag}.

The Euclidean manifold corresponding to this endpoint of the decay is the Euclidean section of the
AdS black hole  with metric
\begin{equation}\label{adsbh}
ds^2 = \left(1+ {r^2 \over R^2} - {M \over C r^2} \right) d\tau^2 + 
\left(1+{r^2 \over R^2} - 
{M \over C r^2}\right)^{-1} \,dr^2  + r^2 \,d\Omega_3^2 + R^2 \, d\Omega_5^2
\,,\end{equation}
where 
 $C=3 {\rm Vol}({\bf S}^3) /16\pi G_{\rm N}$ and $M$ is the mass 
$$
M=C \, \left({r_+^4 \over R^2} +r_+^2 \right).
$$
The horizon radius $r_+$
 is that of the larger of the solutions of the following
 equation
$$
\beta = {2\pi R^2 r_+ \over 2r_+^2 + R^2}\; , 
$$
the smaller solution $r_-$ corresponds 
 to a smaller black hole with negative specific
heat.  In the regime of interest for us, $\beta \ll R$, we have 
 $r_+ \approx  \pi R^2 /\beta$. 

\begin{figure}[ht]
\centerline{\epsfxsize=2.9in\epsfbox{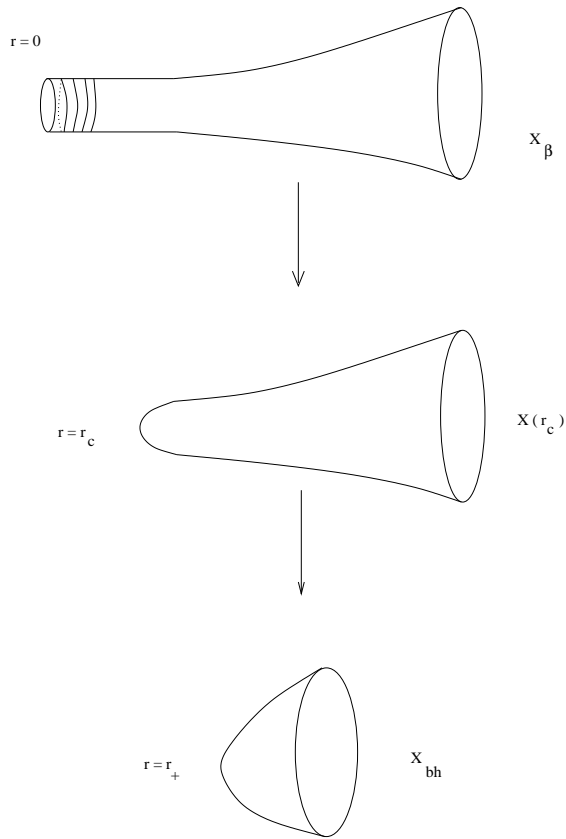}}
\caption{\sl A cartoon of the flow between the original thermal AdS space, with condensing
winding modes in the cylinder region $r\ll R$, and the final Euclidean  AdS black hole,
which does not support winding modes. A set of 
interpolating metrics can be interpreted physically
as off-shell black holes (c.f. \cite{adshag, beken}).  
}
\end{figure}

\begin{figure}[ht]
\centerline{\epsfxsize=2.9in\epsfbox{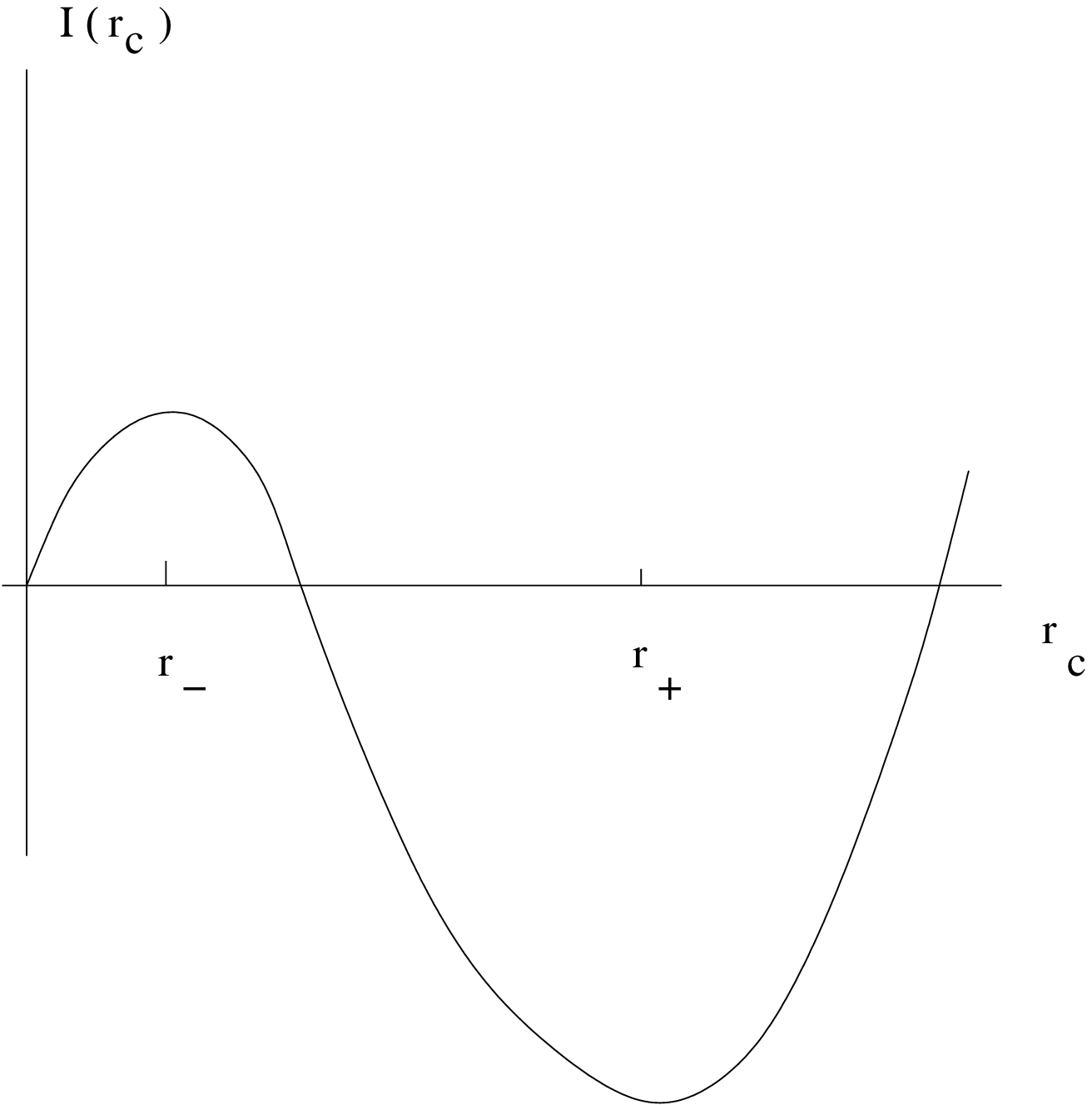}}
\caption{\sl
 The Euclidean action of off-shell black holes with varying mass  and fixed AdS temperature. 
}
\end{figure}

The manifold $X_{\rm bh}$  has topology
${\bf R}^2 \times {\bf S}^3 \times {\bf S}^5$, which can be interpreted as a
 ``capping" of $X_\beta$ that removes the
flat cylinder region $r\ll R$. This capping can be seen as a progressive effect if we consider 
manifolds $X(r_c)$ that interpolate between $X_\beta$ and $X_{\rm bh}$. One such 
set of manifolds  is given by the
Euclidean rotations of off-shell black holes, each of which has a metric
of  the form (\ref{adsbh}), but with
a horizon parameter $r_c$ unrelated to the Euclidean time period $\beta$. In this case, there is a
conical singularity at $r=r_c$ that should be smoothed out 
 by $\alpha'$ corrections in the string theory.  The geometrical
 picture (figure 14)
is very similar to that of non-supersymmetric orbifold decay 
 \cite{aps, minwrev}.

The Euclidean action of these off-shell black holes   reads
\begin{equation}\label{offsa}
I(r_c) = \beta\,M(r_c)- S(r_c) = C  
\,\left[{\beta \,r_c^4 \over R^2 } + \beta \,r_c^2 - {4 \,r_c^3 \over 3} \right]
\end{equation}
and is depicted in figure 15.
 Its shape is very similar to that of the thermal scalar  potential depicted
in figure 13. In fact,
we propose that the description in terms of the thermal scalar condensate
 matches, in the sense of
a string/black-hole correspondence principle, the flow of capped manifolds 
$X(r_c)$ with the value
of the condensate $\bra |\chi|^2 \ket$ 
roughly identified with the extension of the capping, $r_c$. According
to this hypothesis, the function
(\ref{offsa}) would be roughly related to $V_{\rm eff}$ of Eq. (\ref{efaw}) by 
$$
I(r_c) \sim \beta \,R^9\,V_{\rm eff}
\;.$$
This also suggests that the first-order transition of ref. \cite{aw}  
is related to the semiclassical black hole nucleation process studied in
\cite{gpy}. 

One consequence of this line of argument is that, upon removing the infrared regulator
in the original $X_\beta$ manifold, i.e. by sending $R\rightarrow \infty$ at fixed $\beta$ and
fixed $\ell_s$, the stable endpoint manifold $X_{\rm bh}$ recedes to $r_+ \rightarrow \infty$ and
becomes locally flat. Hence, in some sense the endpoint of
 the Hagedorn decay of a hot ten-dimensional space is
a Euclidean, supersymmetric ${\bf R}^{10}$ background (c.f. \cite{adshag}).

Finally, it was  proposed \cite{adshag, beken} that the excluded volume in the
quasi-flat region, ${\rm Vol}_{\rm eff} \sim R^9 - r_c^9$ can be
used as a measure of the number of the 
 depleted degrees of freedom, providing a concept
of ``local central charge" that characterizes the irreversibility of
the renormalization-group flow, analogous to similar concepts in the
theory of boundary flows \cite{boundc}.  

\section{Topology change and winding modes}

\noindent 

We have argued that 
 the $\chi$ condensate is not to be interpreted in terms of ``particles" in the vacuum, but rather as an
order parameter for a topology-change process, in which part of the spacetime is removed by a 
``bubble of nothing" similar to that described in \cite{witb} (see figure 16).  
This relation between tachyon condensation and dynamics of topology change has appeared in other, more
controlled contexts, such as the physics of D-brane annihilation (figure 17)
\cite{bdeath}. The common phenomenon is that a topologically-supported 
string becomes tachyonic. Up to identifications, 
 the string is embedded in the target
space manifold as an interval, i.e. locally 
a copy of ${\bf R}$.
 The condensation can be
envisaged as a process by which the string ``fattens up" into a cylinder by
the transition
\begin{equation}\label{fat}
{\bf R} \rightarrow {\bf S}^{d-1} \times {\bf R}
\,,
\end{equation}
in such a way that 
 the interior of ${\bf S}^{d-1}$ becomes an empty hole in  spacetime.
 Hence, we have a 
wormhole that  grows and ``eats up" the original manifold until infrared effects  
stabilize it (such as negative curvature of AdS in our case). Locally
in Euclidean spacetime,
the flop has the form
\begin{equation}\label{flop}
{\bf R}^d \times {\bf S}^1 \longrightarrow {\bf S}^{d-1} \times {\bf R}^2  
\,.\end{equation}
In this topological jump, a non-contractible ${\bf S}^1$ that  supported the
tachyonic winding modes  becomes contractible, so that the winding modes
can be unwrapped in the new geometry.    
This is entirely similar to the behavior of open strings in the reconnection
process of D-brane annihilation, including the ``fattening" process in
(\ref{fat}) (c.f. figures 16 and 17).

\begin{figure}[ht]
\centerline{\epsfxsize=2.0in\epsfbox{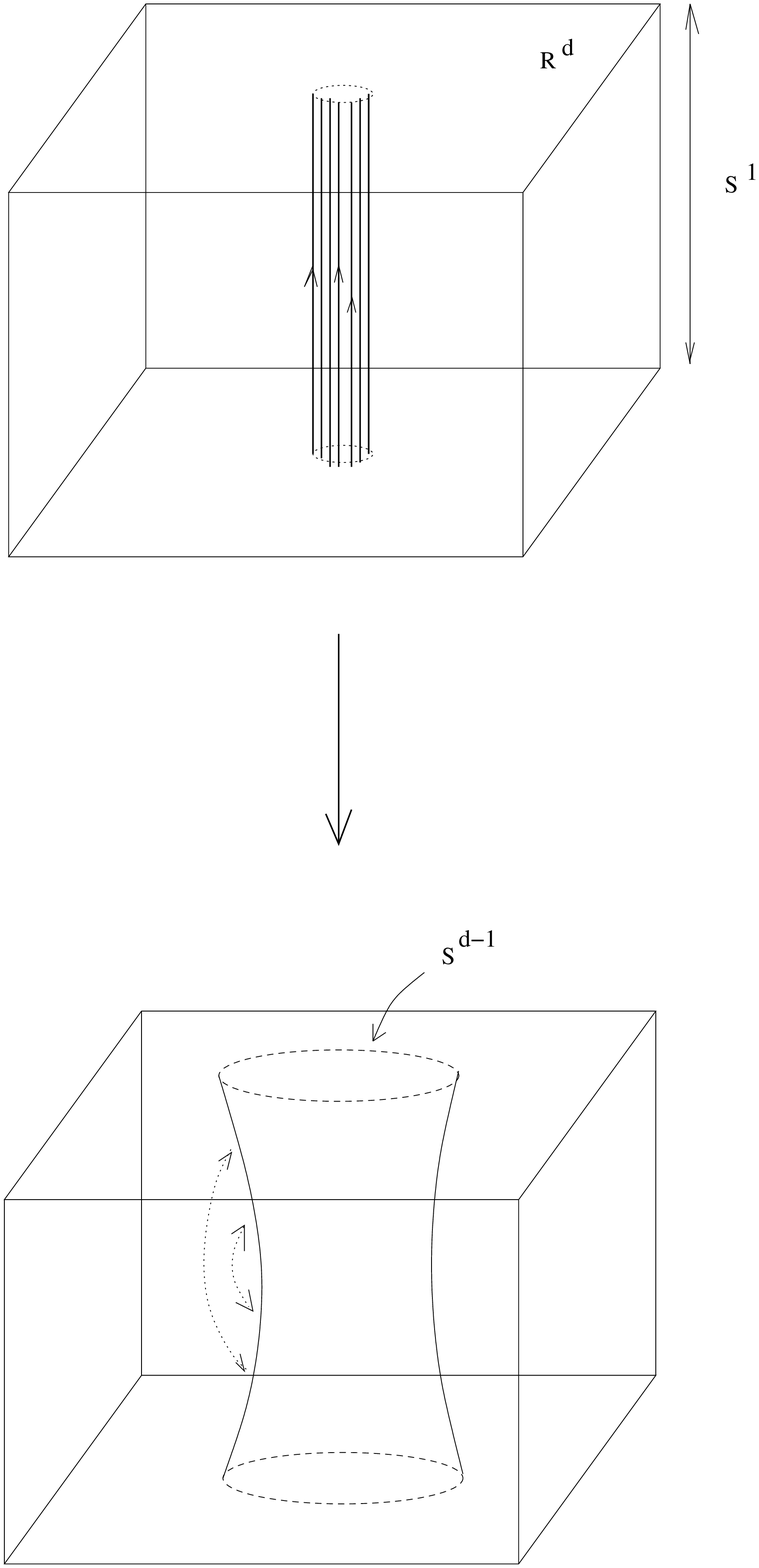}}
\caption{ \sl The bottom picture is 
a local rendering of the basic flop responsible for the
Hagedorn transition: ${\bf R}^d \times {\bf S}^1 \longrightarrow 
{\bf S}^{d-1} \times {\bf R}^2$, that represents the condensation of
many thermal winding strings (top picture). Notice that the ``inside"
of ${\bf S}^{d-1}$ is hollow. In these drawings, opposite points
in the top and bottom planes are identified, together with points
reflected about the plane of symmetry of the throat with ${\bf S}^{d-1}$ 
sections.   
}
\end{figure}

\begin{figure}[ht]
\centerline{\epsfxsize=2.0in\epsfbox{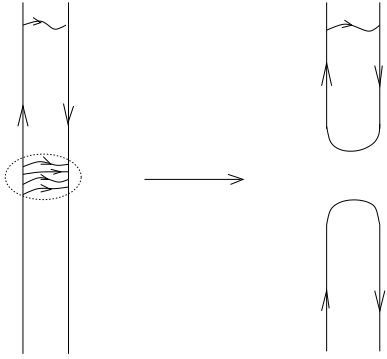}}
\caption{\sl In D-brane annihilation, the condensation of tachyonic strings
stretching between the brane and the antibrane can be depicted semiclassically
as the reconnection of both D$p$-branes. The reconnection generates a throat
that is unstable to growth; this throat is nothing but the original stretched  
string in the ``BIon" picture of ref. \cite{bdeath}, in which the string is locally homeomorphic
to ${\bf R} \times {\bf S}^{p-1}$. 
}
\end{figure}

\subsection{A toy model on supersymmetric cycles} 

\noindent
 
Our geometrical interpretation of the Hagedorn transition is mostly based on
physical considerations in the light of the AdS/CFT correspondence. It would be
interesting to obtain a more explicit derivation of the equivalence between
thermal winding condensation and the topological jump (\ref{flop}). In this respect,
the maximal violation of supersymmetry by the high temperatures is of no particular
help (see, however \cite{sfetsos}). In addition, the winding modes are massless 
at 
$\beta = \beta_s \sim \ell_s $, far from the boundaries of the moduli space of ${\bf S}^1$ compactifications. 
This means that T-duality on the thermal circle ${\bf S}^1_\beta$ (c.f.
\cite{betadual})  is of limited use in elucidating the dynamics involved, since
this dynamics occurs close to the self-dual point. A related system in which these difficulties
can be partially tamed is defined by a variation of the previous AdS/CFT background \cite{thresholds,
martinec, ahanew} (see also \cite{twisted}). 

Consider the type IIB D3-brane background, with the AdS factor in Poincar\'e coordinates and
large $r\rightarrow \infty$ asymptotics 
\begin{equation}\label{dtresm}
ds^2 \longrightarrow
 {r^2 \over R^2} \left(d\tau^2 + dx^2 + d{\vec y}^{\,2} \right) + {R^2 \over r^2} \,dr^2 + 
R^2 \,d\Omega_5^2\;,
\end{equation}
where $\tau$ is compactified with length $\beta$, with supersymmetry-breaking spin structure, whereas
 $x$ is compactified with length $L$, with supersymmetry-preserving spin structure. The coordinates
${\vec y}$ span non-compact ${\bf R}^2$. The dominant background with these boundary conditions is
the black D3-brane in the near-horizon limit, which differs from (\ref{dtresm}) by the
substitutions
$$
d\tau^2 \longrightarrow h(r)\,d\tau^2\;, \qquad dr^2 \longrightarrow dr^2 /h(r)\;,
$$
with $h(r) = 1-(r_0 /r)^4$ and $r_0 = \pi R^2 /\beta$. This background has topology ${\bf R}^2 \times
{\bf R}^2 \times {\bf S}^1 \times {\bf S}^5$, the radial coordinate is restricted to $r\geq r_0$,
 and  it supports winding modes on the circle parametrized
by $x$. In the limit $\beta\rightarrow \infty$, or $r_0 \rightarrow 0$, the winding modes on
circles at small $r$  become massless, and we may ask the question of the what effective dynamics
resolves this singularity. 

In this case, the dynamics of light winding modes at $r\rightarrow 0$ can be transformed into
a more intuitive large-volume effect by standard T-duality on the circles at fixed $r$.  Hence,
at $r\approx r_s = R/L$ the metric is matched to the T-dual
\begin{equation}\label{tdual}
ds^2_{\rm smeared} = {r^2 \over R^2} \left( h(r)\,d\tau^2 + d{\vec y}^{\,2} \,\right) + {R^2 \over r^2} \,
\left({dr^2 \over h(r)}  + d{\tilde x}^{\,2} \,\right) + R^2 \,d\Omega_5^2\;.
\end{equation}
Now ${\tilde x}$ is identified with period $2\pi/L$ and the proper size of the ${\bf S}^1$ fibers
grows as $r\rightarrow 0$. The T-duality transformation allows us to use
the  supergravity of standard momentum modes
 in the T-dual background (\ref{tdual}) to elucidate the dynamics of light
winding modes in the original metric. This T-dual metric is exactly the near-horizon limit
of the so-called ``smeared" D2-brane solution. This is the metric generated by a D2-brane,
averaged over the ${\tilde x}$ coordinate. We can also consider the localized D2-brane solution,
with the same asymptotic conditions, which breaks translational invariance in the ${\tilde x}$
direction and has topology ${\bf R}^2 \times {\bf R}^2 \times {\bf S}^6$. Explicit inspection
shows that the Euclidean action of the localized solution dominates at sufficiently small $r_0$,
leading to a first-order phase transition, since both backgrounds are locally stable at the
transition point (they both have positive specific heat with respect to the temperature $1/\beta$).

Thus, we have a transition on Gregory--Laflamme type \cite{gl}, between locally stable backgrounds,
with a basic topological jump given by
the flop
$$
{\bf S}^1 \times {\bf S}^5 \longrightarrow {\bf S}^6\;.
$$
We see that the non-contractible cycle
 supporting the original winding modes again disappears
and is replaced by a contractible geometry, along the lines of our general scenario in the
last section. We can think of the transition as the result of winding-mode condensation, because
the breaking of translational invariance in ${\tilde x}$ is equivalent to the condensation of
momentum modes in the background (\ref{tdual}), which are exactly  the winding modes of the
original vanishing cycle.    

\begin{figure}[ht]
\centerline{\epsfxsize=3.0in\epsfbox{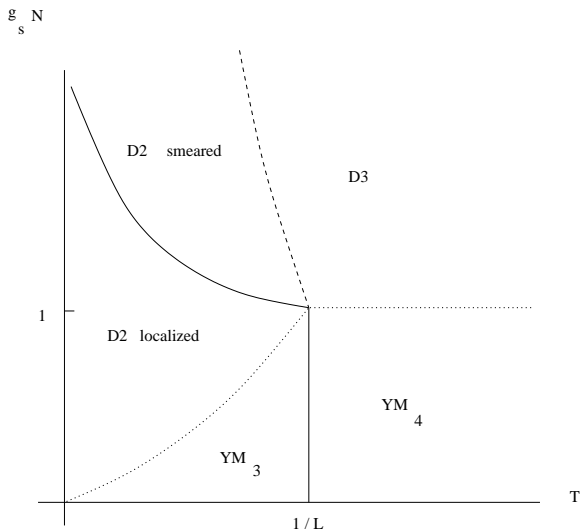}}
\caption{\sl Phase diagram of the D3-brane theory as a function of the 't Hooft coupling $g_s N$
and the temperature, with
 a compactified direction of length $L$ (c.f. \cite{thresholds}). 
At weak coupling $g_s N <1$ we have the Yang--Mills gas descriptions in
four or three dimensions, with the standard threshold for finite size effects at $LT \sim 1$.
At strong coupling, $g_s N >1$ we have the AdS dual description given by the black D3-brane metric
at high temperatures. At lower temperatures we have a smooth transition to the smeared black D2-brane
metric and then a topological transition to the localized black D2-brane metric. The dotted lines
denote correspondence lines between perturbative Yang--Mills descriptions and supergravity descriptions,
whereas full lines denote thresholds for finite-size effects on the thermodynamic functions. 
 The localization transition is of first order in the supergravity
approximation. The thermodynamic functions calculated in the classical gravity approximation
 do not change across the T-duality transition  (dashed line). The supergravity backgrounds
to the left of the T-duality transition are dominated by modes that are interpreted as light
winding modes form the point of view of the original D3-brane metric. More general phase
diagrams, including effects of large-$N$ nonlinearities appear in \cite{hoy}.   
 }
\end{figure}

The interpretation of these geometrical transitions in the dual gauge theory is interesting. They
are large-$N$ phase transitions that characterize the effect of a finite-size toroidal compactification
in the thermodynamics of the gauge theory. At weak coupling, this is just the change in the
scaling of the entropy  from $S\sim T^3$ to $S\sim T^2$ as the temperature drops below the
threshold $T_L \sim 1/L$ (see figure 18).  

\section{Conclusion}

\noindent

We have discussed some, but by no means all, systems, which have a Hagedorn
ridge.
What was common to all the systems discussed was that for them the ridge is just
a transient one passes as one increases their energy.  We have described
various components which each play in their turn  a role in defining the
thermodynamical properties of strings.  In particular a variety of  black holes
and black strings  turn out to best represent the degrees of freedom at very
high energy.

  We have discussed both broad and detailed features of the Hagedorn
ridge.  Ian's work and many following ones imagine that spacetime melts away
as one reaches the end point of the ridge.  The separation into perturbative
world-sheet physics (i.e. a string picture) and target space (i.e. a geometry
in which the strings propagate) becomes questionable.
While these are perhaps yet to be uncovered phases of gravity, their properties,
as consistent world-sheet theories possessing a total vanishing Virasoro central
charge, are yet to be elaborated on. 

In this essay we brought up a milder form
of the disappearance as well as topological transmutation of chunks of spacetime 
 but not all of it. The nucleation of black holes has become important at
the edge of the Hagedorn ridge. Portions of space-time have been expelled as
these black holes stabilize, they cause a topology change and in some cases
even lead all the way to flat ten dimensional supersymmetric space-time.  There
may yet be vistas to be discovered and bumps to be negotiated as one
traverses the ridge. We miss Ian's guidance for these future ventures.

\section*{Acknowledgments}

We would like to thank Steve Abel, Ofer Aharony,
Shmuel Elitzur and Miguel A. V\'aquez--Mozo for many
discussions and/or  collaborations on the subject of this essay. 
 The work of J.L.F.B. was partially supported by MCyT
 and FEDER under grant
BFM2002-03881 and
 the European RTN network
 HPRN-CT-2002-00325. The work of E.R. is supported in part by the
Miller Foundation, the
BSF-American Israeli Bi-National Science Foundation, The Israel Science
Foundation-Centers of Excellence Program, The German-Israel Bi-National
Science Foundation and the European RTN network HPRN-CT-2000-00122.


\newpage
\begin{thebibliography}{99}

\bibitem{hag}
R. Hagedorn, {Suppl.~Nuovo Cimento} {\bf 3}, 147 (1965).

\bibitem{ian}
I.I. Kogan, {\it JETP Lett.} {\bf 45}, 709 (1987); B. Sathiapalan,
\prd{35}{3277}{1987}.



\bibitem{cab}
N. Cabibbo and G. Parisi, \plb{59}{67}{1975}.

\bibitem{thorn}
C. Thorn, \plb{99}{458}{1981}.


\bibitem{bunch}
S. Fubini and G. Veneziano, { Nuovo Cimento A} {\bf 64}, 1640 (1969);
K. Huang and S.~Weinberg, \prl{25}{895}{1970};
S. Frautschi, \prd{3}{2821}{1971};
R.D. Carlitz, \prd{5}{3231}{1972};
E. Alvarez, \prd{31}{418}{1985}; \npb{269}{596}{1986};
M. Bowick and L.C.R. Wijewardhana, \prl{54}{2485}{1985};
B. Sundborg, \npb{254}{883}{1985};
S.N. Tye, \plb{158}{388}{1985};
E. Alvarez and M.A.R. Osorio, \prd{36}{1175}{1987};
P. Salomonson and B. Skagerstam, \npb{268}{349}{1986}; {Physica A}
{\bf 158}, 499 (19890;
D. Mitchell and N. Turok, \prl{58}{1577}{1987}; \npb{294}{1138}{1987};
I. Antoniadis, J. Ellis and D.V.~Nanopoulos, \plb{199}{402}{1987};
M. Axenides, S.D. Ellis and C. Kounnas, \prd{37}{2964}{1988};
M. Mc Guigan, \prd{38}{552}{1988};
A.A. Abrikosov Jr. and Ya. I. Kogan
\ijmpa{6}{1501}{1991} (submitted 1989),
{Sov. Phys. JETP} {\bf 69}, 235 (1989);
M.J. Bowick and S.B. Giddings, \npb{325}{631}{1989};
S.B. Giddings, \plb{226}{55}{1989};
F. Englert and J. Orloff, \npb{334}{472}{1990};
S.A. Abel, \npb{372}{189}{1992};
J.L.F. Barb\'on and M.A. V\'azquez-Mozo,
\npb{497}{236}{1997} \hepth{9701142};
M. Laucelli Meana, M.A.R.~Osorio and J. Puente Penalba, \plb{400}{275}{1997} \hepth{9701122};
\plb{408}{183}{1997} \hepth{9705185};
S.S. Gubser, S. Gukov, I.R. Klebanov, M. Rangamani and E. Witten,
{J. Math. Phys.} {\bf 42}, 2749 (2001) \hepth{0009140};
J.L.F. Barb\'on and E. Rabinovici, {JHEP} {\bf 06}, 029 (2001)
\hepth{0104169};
K. R. Dienes and M. Lennek, {\em Thermal duality confronts entropy:
a new approach to string thermodynamics?} 
hep-th/0312173;
 M.A.~Cobas, M.A.R. Osorio and  M. Suarez,
{\em Thermodynamic nonextensivity in a closed string gas},
hep-th/0406043;
S. Chaudhuri, {\em Thermal Duality and the Canonical String Ensemble},
hep-th/0409301.


 
\bibitem{banks}
G. 't Hooft, \npb{256}{727}{1985};
T. Banks, {\em A critique of pure string theory: Heterodox opinions of diverse dimensions}, hep-th/0306074.

\bibitem{usabel}
S.A. Abel, J.L.F. Barb\'on, I.I. Kogan and E. Rabinovici, {JHEP} {\bf 04},
 015 (1999)\\{} \hepth{9902058}.

\bibitem{golfand} 
S.A. Abel, J.L.F. Barb\'on, I.I. Kogan and E. Rabinovici, 
{\em Some thermodynamical aspects of string theory},
hep-th/9911004.

\bibitem{holo}
G. 't Hooft, 
{\em Dimensional Reduction In Quantum Gravity},'' gr-qc/9310026;\\{}
J. Bekenstein, \prd{49}{1912}{1994} [gr-qc/9307035];\\{}
L. Susskind, {J. Math. Phys.} {\bf 36}, 6377 (1995) \hepth{9409089}.

\bibitem{lst} 
O. Aharony, M. Berkooz, D. Kutasov and N. Seiberg, {JHEP} {\bf 10}, 004 (1998)\\{}
\hepth{9808149}; O. Aharony, {Class. Quant. Grav.} {\bf 17}, 929 (2000) 
\hepth{9911147;} D. Kutasov, 
Review prepared for ICTP Spring School on Superstrings and Related Matters,
Trieste, Italy, 2-10 Apr 2001.
Published in *Trieste 2001, Superstrings and related matters* 165-209.  

\bibitem{bmn}
D. Berenstein, J.M. Maldacena and H. Nastase, {JHEP} {\bf 0204}, 013 (2002)\\{} 
\hepth{0202021}.

\bibitem{kutsaha}
D. Kutasov and V. Sahakyan, 
JHEP {\bf 0102}, 021 (2001)
[hep-th/0012258].

\bibitem{bmnt}
L.A. Pando-Zayas and D. Vaman, \prd{67}{106006}{2003} \hepth{0208066};
B.R. Greene, K. Schalm and G. Shiu, \npb{652}{105}{2003} \hepth{0208163};
R.C. Brower, D.A. Lowe and C-I. Tan, \npb{652}{127}{2003}[hep-th/\-0211201];
Y. Sugawara, \npb{661}{191}{2003} \hepth{0301035};
G.~Grignani, M. Orselli, G.W. Semenoff and D. Trancadelli, {JHEP} {\bf 0306},
006 (2003) \hepth{0301186};
F.~Bigazzi and   A.L. Cotrone,
{JHEP} {\bf 0308} 052 (2003)
\hepth{0306102};
R. Apreda, F. Bigazzi and A.L. Cotrone,
{JHEP} {\bf 0312} 042 (2003)
\hepth{0307055}.




\bibitem{branvafa}
R. Brandenberger and C. Vafa, \npb{316}{391}{1989}.

\bibitem{obv}
R. Brandenberger, D. A. Easson and D. Kimberly,
\npb{623}{421}{2002}
\hepth{0109165}; 
S. A. Abel,  K. Freese and I. I. Kogan,
{JHEP} {\bf 0101} 039 (2001) 
 \hepth{0005028}.  

\bibitem{deo}
N. Deo, S. Jain and C.-I. Tan, \plb{220}{125}{1989};
\prd{40}{2646}{1989}; N. Deo, S. Jain, O. Narayan and C.-I. Tan,
\prd{45}{3641.}{1992}.

\bibitem{love}
D.A. Lowe and L. Thorlacius, \prd{51}{665}{1995},  \hepth{9408134};\\{}
S. Lee and L. Thorlacius, \plb{413}{303}{1997}, \hepth{9707167}.

\bibitem{brab}
T. Banks and E. Rabinovici, \npb{160}{349}{1979}; \\{}  
D. Forster, \plb{77}{211}{1978}; \\{}
M. Stone and P. Thomas, \prl{41}{351}{1978}. 

\bibitem{boundc}
I. Affleck and A.W.W. Ludwig, \prd{67}{161}{1991};
S. Elitzur, E. Rabinovici and G. Sarkisian, \npb{541}{246}{1999}
\hepth{9807161};
J.A. Harvey, D.~Kutasov and E.J. Martinec, 
{\em On the relevance of tachyons},
hep-th/0003101.

\bibitem{bdeath}
C. G. Callan,Jr and  J. M. Maldacena,
\npb{513}{198}{1998}
\hepth{9708147}.
 
\bibitem{mav}
M.B. Green \plb{266}{325;}{1991} \plb{329}{435}{1994} \hepth{9403040};
M.A. V\'azquez-Mozo, \plb{388}{494,}{1996} \hepth{9607052}.

\bibitem{thresholds}
J.L.F. Barb\'on, I.I. Kogan and E. Rabinovici, \npb{544}{104}{1999}\\{}
 \hepth{9809033}.

\bibitem{aw}
J. Atick and E. Witten, \npb{310}{291}{1988}.

\bibitem{gpy}
D.J. Gross, M.J. Perry and
L.G. Yaffe,
\prd{25}{330}{1982}.


\bibitem{hp}
G.T. Horowitz and J. Polchinski, \prd{55}{6189}{1997} \hepth{9612146}.

\bibitem{magoo}
O. Aharony, S. S. Gubser, J. M. Maldacena, H. Ooguri and Y. Oz,
{Phys. Rept.} {\bf 323}, 183 (2000) 
\hepth{9905111.}
 
\bibitem{adscft}
J.M.  Maldacena, {Adv. Theor. Math. Phys.} {\bf 2}, 231 (1998)
\hepth{9711200};
S.S.~Gubser, I.R. Klebanov and A.M. Polyakov,
\plb{428}{105}{1998}
\hepth{9802109};
E. Witten,
{Adv. Theor. Math. Phys.} {\bf 2}, 253 (1998) \hepth{9802150}.

\bibitem{corr}
G. Veneziano, {Europhys. Lett.} {\bf 2}, 199 (1986);
G. Veneziano, in  {\em Hagedorn Festschrift.}
Eds: Jean Letessier, Hans Gutbrod and Johann Rafelski,
NATO-ASI Series B: Physics {\bf 346}, 63 ( Plenum Press, New York, 1995);
L. Susskind, 
{\em Some Speculations About Black Hole Entropy In String Theory},
hep-th/9309145.

\bibitem{cor}
G.T. Horowitz and J. Polchinski, \prd{57}{2557}{1998} \hepth{9707170};\\{}
T. Damour and G. Veneziano, \npb{568}{93}{2000} \hepth{9907030}.  

\bibitem{witb}
E. Witten, \npb{195}{650;}{1982}
S.P. de Alwis and A.T. Flournoy, \prd{66}{026005}{2002} \hepth{0201185};
O. Aharony, M. Fabinger, G.T.~Horowitz and E. Silverstein, {JHEP} {\bf 0207}, 
007 (2002) \hepth{0204158}.

\bibitem{beken}
J.L.F. Barb\'on and E. Rabinovici, {Found. Phys.} {\bf 33}, 145 (2003)
\hepth{0211212.}

\bibitem{aps}
A. Adams, J. Polchinski and E. Silverstein, {JHEP} {\bf 0110}, 029 (2001)
\hepth{0108075};
J.A. Harvey, D. Kutasov, E.J. Martinec and G. Moore, 
{\em Localized tachyons and RG flows},
hep-th/0111154.

\bibitem{sfetsos}
I. Antoniadis and C. Kounnas, \plb{261}{369}{1991};
I. Antoniadis, J.P.~Derendinger and C. Kounnas, 
Nucl.\ Phys.\ B {\bf 551}, 41 (1999)
[hep-th/9902032];
I. Bakas, A. Bilal, J.P. Derendinger and K. Sfetsos, 
Nucl.\ Phys.\ B {\bf 593}, 31 (2001)
[hep-th/0006222].

\bibitem{aha}
B. Sundborg, 
Nucl.\ Phys.\ B {\bf 573}, 349 (2000)
[hep-th/9908001];
O. Aharony, J.~Marsano, S. Minwalla, K. Papadodimas and M. Van Raamsdonk,
{\em The Hagedorn / deconfinement phase transition in weakly coupled large N gauge theories}, hep-th/0310285.


\bibitem{betadual}
R. Rohm, \npb{237}{553}{1984}; 
E. Alvarez and M.A.R. Osorio, \prd{40}{1150}{1989};
O. Bergman and M.R. Gaberdiel, {JHEP} {\bf 07}, 022 (1999) \hepth{9906055}.

\bibitem{roth}
B. McClain and B. Roth, {Commun. Math. Phys.} {\bf 111}, 539  (1987);\\{}
E. Alvarez and M.A.R. Osorio, \npb{304}{327}{1988}.  

\bibitem{minwrev}
M. Headrick, S. Minwalla and T. Takayanagi,
Class.\ Quant.\ Grav.\  {\bf 21}, S1539 (2004)
[hep-th/0405064].

\bibitem{sen}
A. Sen, {\em Non-BPS states and branes in string theory},
hep-th/9904207;
JHEP {\bf 9808}, 012 (1998)
[hep-th/9805170].

\bibitem{polc}
J. Polchinski, {Commun. Math. Phys.} {\bf 104}, 37 (1986).

\bibitem{mavoso}
M.A.R. Osorio and M.A. V\'azquez-Mozo, 
\plb{280}{21}{1992}\\{} \hepth{9201044}; \prd{47}{3411}{1993} \hepth{9207002}.

\bibitem{martinec}
M. Li, E. Martinec and V. Sahakyan, 
Phys.\ Rev.\ D {\bf 59}, 044035 (1999)\\{}
[hep-th/9809061];
E. Martinec and V. Sahakyan, 
Phys.\ Rev.\ D {\bf 59}, 124005 (1999)
[hep-th/9810224];
{\em ibid} {\bf 60}, 064002 (1999)
[hep-th/9901135].

\bibitem{ahanew}
O. Aharony, J. Marsano, S. Minwalla and T. Wiseman, 
{\em Black hole - black string phase transitions in thermal 1+1 dimensional supersymmetric Yang-Mills theory on a circle},
hep-th/0406210.
T. Harmark and  N. A. Obers,
{\em New Phases of Near-Extremal Branes on a Circle},  hep-th/0407094.



\bibitem{twisted}
 J. R. David, M. Gutperle, M. Headrick and  S. Minwalla,
{JHEP} {\bf  0202}, 041 (2002) 
\hepth{0111212}.
 
\bibitem{HP}
S.W. Hawking and D. Page, 
Commun.\ Math.\ Phys.\  {\bf 87}, 577 (1983).

\bibitem{w}
E. Witten, {Adv. Theor. Math. Phys.} {\bf 2}, 505 (1998)
\hepth{9803131};
T. Banks, M.R.~Douglas, G.T. Horowitz and E. Martinec, 
{\em AdS dynamics from conformal field theory},
hep-th/9808016;
J.L.F. Barb\'on and E. Rabinovici,
\npb{545}{371}{1999} \hepth{9805143}.

\bibitem{gl}
R. Gregory and R. Laflamme, \prl{70}{2837}{1993} \hepth{9301052};
\npb{428}{399}{1994} \hepth{9404071}; \prd{51}{305}{1995}\\{} \hepth{9410050};
B. Kol,  {\em Topology Change in General Relativity and the Black Hole Black String Transition},
 hep-th/0206220.

   

\bibitem{adshag}
J.L.F. Barb\'on and E. Rabinovici, 
{JHEP} {\bf 0203}, 057 (2002) \hepth{0112173}. 

\bibitem{hoy}

J.L.F. Barb\'on and C. Hoyos, 
{JHEP} {\bf 0401}, 049 (2004) 
\hepth{0311274}.  


\end{thebibliography}
\end{document}